\documentstyle[12pt]{article}

\def\k{{k_F}}
\def\intt#1{\int \frac{d^3#1}{(2\pi)^3}}
\def\bfm#1{{\mbox{\boldmath $#1$}}}
\def\ssi{\sum_{\scriptscriptstyle \rm spin\atop \scriptscriptstyle \rm isospin}}
\def\MA{(M^0_{A-1})^2}
\def\E{{\cal E}}
\def\pslash{p\!\!/}
\begin{document}
\baselineskip=1.4\baselineskip
\pagestyle{empty}
\begin{titlepage}
\title{Inclusive versus Exclusive EM Processes \\ in Relativistic Nuclear
Systems
\footnote[1]{This work is supported in part by funds provided by the U.S.
Department of Energy (D.O.E.) under cooperative agreement
\#DE-FC01-94ER40818.
\vskip0.5truecm
MIT/CTP \#2523}
\footnote[2]{Submitted to {\it Phys. Rev. C}}
}
\author{R.Cenni$^\dagger$, T.W.Donnelly$^\ddagger$ and A.Molinari$^\sharp$
\\~ \\$^\dagger$Dipartimento di Fisica -- Universit\`a di Genova\\
Istituto Nazionale di Fisica Nucleare -- sez. di Genova\\
Via Dodecaneso, 33 -- 16146 Genova (Italy)\\~
\\$^\ddagger$ Center for Theoretical Physics \\
Laboratory for Nuclear Science and Department of Physics\\
Massachusetts Institute of Technology\\
Cambridge, Massachusetts 02139, U. S. A.\\~
\\ $^\sharp$ Dipartimento di Fisica Teorica dell'Universit\`a di Torino
\\
Istituto Nazionale di Fisica Nucleare --  Sez. di Torino\\
Via P. Giuria, 1 -- 10125  Torino (Italy)}
\date{14 June 1996}
\maketitle
\end{titlepage}
\pagestyle{plain}
\begin{abstract}
\vskip2cm
{\bf Abstract:}
Connections are explored between exclusive and
inclusive electron scattering within the framework of the 
relativistic plane-wave impulse approximation, beginning with an analysis of
the model-independent kinematical constraints to be found in the 
missing energy--missing momentum plane. From the interplay between these 
constraints and the spectral function basic features of the exclusive and 
inclusive nuclear responses are seen to arise.
In particular, the responses of the relativistic Fermi gas
and of a specific hybrid model with confined nucleons in 
the initial state are compared in this work.
As expected, the exclusive responses are significantly different in
the two models, whereas the inclusive ones are rather similar. 
By extending previous work on the relativistic Fermi gas, 
a reduced response is introduced for the hybrid model such that it
fulfills the Coulomb and the higher-power energy-weighted sum rules. While 
incorporating specific classes of off-shellness for the struck nucleons, it 
is found that the reducing factor required is largely model-independent and, 
as such, yields a reduced response that is useful for extracting the 
Coulomb sum rule from experimental data.
Finally, guided by the difference between the energy-weighted sum rules of the
two models, a version of the relativistic Fermi gas is devised which has the
0$^{\rm th}$, 1$^{\rm st}$ and 2$^{\rm nd}$ moments of the charge response
which agree rather well with those of the hybrid model: this version thus
incorporates {\em a priori} the binding and confinement effects of the 
stuck nucleons while retaining the simplicity of the original Fermi gas.
\end{abstract}
\newpage

\section{Introduction\label{sect1}}

The relativistic non-interacting Fermi gas (RFG) is a rare example
of a three-dimensional many-body system that can be solved while fully
respecting Lorentz covariance. When used in interpreting inclusive electron 
scattering at energies and momentum transfers that are not too small 
({\it i.e.,} at the scales set by the Fermi momentum) it leads 
to results that qualitatively account for the existing data. 
It thus seems natural to ask the question: for exclusive processes, {\it e.g.}
for the coincidence experiment ($e, e^\prime p$), is the RFG
able to account for experiment as well? The answer to this question is  
clearly negative.

One might then argue that the spectral function of the RFG, the
key ingredient in obtaining the exclusive cross section in the
framework of the Plane-Wave Impulse Approximation (PWIA), is severely out of 
touch with the physical world. Yet when integrated over the energies and 
directions of the outgoing nucleons it leads to sensible results.
In other words the RFG provides an acceptable first approximation 
to the particle-hole Green's
function, which is closely related to the inclusive cross section, but not 
to the single-particle propagator, which is closely related to the spectral 
function.

Of course, some of the failure of the RFG to explain such exclusive processes 
also, but clearly not only, arises from shortcomings shared with the PWIA
in dealing with the $(e, e^\prime p)$ reaction through neglect of 
final-state interactions. In this work we limit our focus to kinematics 
where such effects are believed to be small and accordingly where the 
PWIA should provide a good starting point.

With this as background, let us now state the motivations for the present 
work: we explore within the framework of the PWIA the connection between
exclusive and inclusive EM processes with the aim of
\begin{enumerate}

\item deepening the understanding of the RFG model;
\item assessing the impact of confinement effects for struck nucleons in 
a finite nucleus on the inclusive charge response function by
exploring a specific hybrid model (to be defined in detail later) 
whose simple, tractable form facilitates taking the limit $A\to\infty$;
\item providing a link between finite and infinite Fermi systems
in searching  for a common {\it model-independent} factor to be used in 
defining a reduced charge response such as to have the Coulomb sum rule 
fulfilled and
\item exploring whether a modified RFG charge response can be worked out in
such a way that, on the average, incorporates some of the properties of a 
confined system described in a mean-field framework.
\end {enumerate}

To start with in sec.~\ref{sect2} and \ref{sect3}
the kinematics of exclusive processes
and the PWIA are shortly revisited together with the definitions and
basic properties of the scaling variable and spectral function.
In sec.~\ref{sect4} the inclusive cross sections and 
associated nuclear responses are obtained by
suitably integrating over the exclusive processes.
In particular we pay attention 
to the problem of fixing a dividing factor to yield a reduced 
longitudinal response for a finite system such that it obeys
the Coulomb sum rule.

Next, in sec.~\ref{sect5}, we examine in some depth the RFG, including the 
associated spectral
function and its support in the plane of the momentum and excitation
energy of the daughter nucleus (often referred to as ``missing momentum'' 
and ``missing energy'', respectively). In particular,
by integrating the exclusive process over the variables of the 
outgoing nucleon, we recover the well-known expression for the
inclusive cross section of the RFG. This we achieve for 
$q \geq 2 k_F$ ($q$ being the three-momentum transfer and
$k_F$ the Fermi momentum), but not for $q <2 k_F$ where Pauli 
correlations play a role. Indeed the PWIA ignores all correlations between 
the outgoing nucleon and the daughter nucleus, including those stemming 
from the fermion statistics.

In sec.~\ref{sect6} we improve upon  the spectral function 
of the RFG allowing the 
hole states to be bound by a harmonic oscillator potential, the particle states
still being plane waves. Thus we set up a simple, tractable hybrid model (HM)
to guide us in exploring the connections between finite and infinite Fermi 
systems. Our goal here is not to undertake computations with the best 
mean-field wave functions (which of course could be done using the ideas 
presented in this work), but to develop a model for the initial nuclear
state where the nucleons are confined, on the one hand, and yet simply 
described, on the other. For such a model much of the analysis can be done 
analytically and it is feasible to consider very heavy nuclei: in 
sec.~\ref{sect6} we present results for closed major shell $N=Z$ 
nuclei up to the (hypothetical) case $A=336$. The harmonic oscillator HM
represents an extreme when compared with the RFG in that the former has 
very strongly confined bound-state wave functions (with gaussian tails 
rather than exponentials as would be the case when finite potentials are 
used), whereas the latter only involves plane waves.

In inter-comparing the two models we find that the Fermi momentum $k_F$ of the
RFG cannot be adjusted to yield the correct root-mean-square momentum for each
shell of a harmonic oscillator well of a given frequency
$\omega_0$. To achieve this a set of RFG, each with its own $k_F$, is needed. 
When dealing with inclusive processes, however, 
one global $\k$ is usually selected to get the same response width 
as that produced in some other model such as the HM. These alternatives 
are also discussed in sec.~\ref{sect6}.

In the same section we continue with an assessment of 
the impact on the longitudinal response of the finite size of the 
bound nuclear system in a mean field--PWIA framework.
We find that the positive separation energy (but not the confinement) of 
the nucleus induces a hardening of the longitudinal
response with respect to the RFG result. On the basis of
scaling arguments we quantify this energy shift through a simple, but useful 
expression for the difference between the HM and RFG energy-weighted sum 
rules. We search for a Fermi momentum $k_F$ and for a shift of the excitation
energy such as to account for the confinement and separation energy of the 
nucleons in the initial state when calculating the longitudinal 
response within the framework of the RFG model. We accomplish this 
by comparing the inclusive charge response of the HM and of the
RFG together with their various moments (the sum rule, mean 
excitation energy and variance). 

We succeed as well in solving the problem of 
finding a reduction factor for the charge response in a way that permits the 
single-nucleon physics to be factored out and the Coulomb sum rule to be 
satisfied asymptotically. In this connection we extend
a result found previously for the RFG. Namely, 
the reduction factor devised for the RFG is valid as well for the
HM with harmonic oscillator wave functions and, by extension, suggests that 
such a definition should prove to be useful for wide classes of mean-field 
potentials.

Finally, our conclusions are presented in sec.~\ref{sect7}.

\section{The kinematics\label{sect2}}

\begin{figure}[h]
\vskip8cm
\includegraphics{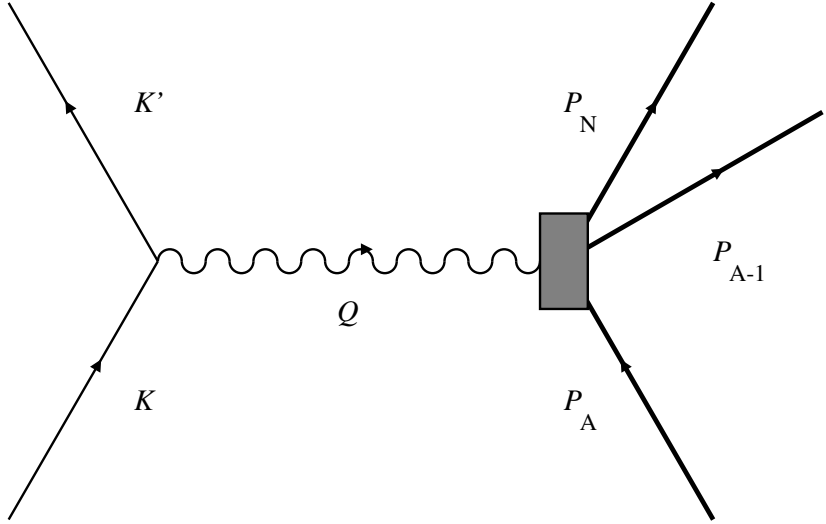}
\caption{\protect\label{Fig1} The diagram describing the coincidence 
process $(e,e^\prime p)$.}
\end{figure}

In this section we consider exclusive (semi-inclusive) nuclear
electron scattering where the final electron is detected in coincidence
with an outgoing nucleon, as described by the diagram of 
fig. \ref{Fig1}  in the one-photon-exchange approximation. The associated 
cross section (in the laboratory system where the target four-momentum is 
$P^\mu=(M_A^0,0)$, with $M_A^0$ the initial ground-state mass, and 
differential with respect to both the outgoing electron and nucleon variables) 
reads \cite{Raskin89}
\begin{eqnarray}
\lefteqn{
{{d^4\sigma} \over {d\varepsilon^\prime d\Omega_e dE_Nd\Omega_N}}
= {{4\alpha^2} \over {(2\pi)^3}} {{m^2_e M_{A-1}m_N p_N k^\prime}
\over {\varepsilon E_{A-1}}} }
\label{II.1}\\
&&\times {1\over{Q^4}} \eta_{\mu\nu} W^{\mu\nu} \delta 
\left(\omega+M^0_A-E_N-E_{A-1}\right).
\nonumber
\end{eqnarray}
In the above the square of the (spacelike) four-momentum transfer is given by 
$Q^2=\omega^2-q^2<0$, where $Q^\mu=(\omega, {\bf q})$,
\begin{equation}
\eta_{\mu \nu}\ 
={1\over 2m_e^2}\left[\ K_\mu K^\prime_\nu + K^\prime_\mu K_\nu - g_{\mu\nu}
K \cdot K^\prime\right]
\label{II.1bis}
\end {equation}
is the symmetric ultrarelativistic leptonic tensor and $W^{\mu\nu}$ its
hadronic counterpart whose expression will be discussed later. Furthermore,
${\bf p}_N$ and $E_N=(m_N^2+p_N^2)^{1/2}$ are the momentum and energy of the 
ejected nucleon, {\it i.e.,} $P_N^\mu=(E_N, {\bf p}_N)$.
The other variables are either self-explanatory or 
similar to those used in ref. \cite{Cab93} (recall also that 
$e^2 = 4 \pi \alpha$).

In deriving eq.~(\ref{II.1}) the conventions of Bjorken and 
Drell \cite{BjDr-64-B} have been used.
As a consequence the single-nucleon states are normalized according to
\begin{equation}
<{\bf p}|{\bf q}>=(2\pi)^3{E({\bf p})\over m_N}\delta({\bf p}-{\bf q}),
\label{II.1.1.1}
\end{equation}
with $E({\bf p})=(m_N^2+p^2)^{1/2}$, 
which is consistent with the anticommutation rule
\begin{equation}
\{\hat a^\dagger_p, \hat a_q\}=(2\pi)^3{E({\bf p})\over m_N}\delta({\bf p}-
{\bf q})
\label{II.1.2}
\end{equation}
for the fermion operators.
For the initial ($<A|$) and daughter ($<A-1|$) unfragmented nuclear 
states we use the normalization 
\begin{equation}
<A|A>={E_A^0\over M^0_A}=1,\qquad\qquad <A-1|A-1>={E_{A-1}\over M_{A-1}}
\label{II.1.3}
\end{equation}
(throughout we work in the laboratory system where the initial nucleus is at 
rest). Note that when the state $|A-1>$ is fragmented its normalization is
given by the product of the normalizations of the fragments and in such
a case eq. (\ref{II.1}) should be modified accordingly.

Let us now discuss the energy-conserving $\delta$-function in
eq.~(\ref{II.1}); for this purpose it helps to introduce, following ref.
\cite{DaMcDoSi-90}, the non-negative variable
\begin{equation}
{\cal E}\ =\ E_{A-1} - E^0_{A-1} = {\sqrt{{\bf  p}^2_{A-1} + 
(M_{A-1})^2}}\ -\ {\sqrt{{\bf p}^2_{A-1} + \MA}},
\end{equation}
which represents the excitation energy of the residual $A-1$ nucleus
moving with momentum ${\bf p}_{A-1}$. Here $E_{A-1}$ is the energy of the 
daughter system when it is generally excited, while $E_{A-1}^0$ is the 
energy when in its ground state. 
In the laboratory frame, where ${\bf p}_A = 0$, one has necessarily that
${\bf p}_{A-1} = - {\bf p}$, if ${\bf p}_{N} = {\bf p} + {\bf q}$ is the
momentum of the outgoing nucleon. The argument of the $\delta$-function 
then vanishes for
\begin{equation}
\omega\ =\  {\cal E}+ E_S + \left( {\sqrt{({\bf p} + {\bf q})^2 + m_N^2}}
\ -\ m_N \right) + \left( {\sqrt{{\bf p}^2 + \MA}} - M^0_{A-1} \right),
\label{XX0}
\end{equation}
where (for a stable nucleus) the non-negative separation energy
\begin {equation}
E_S = M_{A-1}^0 + m_N - M^0_A \geq 0
\label{XX00}
\end{equation}
has been introduced.

Another energy that will prove to be relevant in our later discussions is 
$E\equiv E_N-\omega$, that is, we have 
$P^\mu\equiv P_N^\mu-Q^\mu=(E, {\bf p})$. In general $P^2=E^2-p^2\neq m_N^2$ 
and thus the kinematics for this four-vector are not those of an 
on-shell nucleon. Defining the daughter ground-state recoil kinetic energy 
$T_{A-1}^0\equiv E_{A-1}^0-M_{A-1}^0$ we may use eq.~(\ref{XX0}) to write
\begin{equation}
E=m_N-({\cal E}+E_S+T_{A-1}^0).
\label{XX1}
\end{equation}
This may then be compared with the energy of an on-shell nucleon having 
three-momentum $p$, namely, using the nomenclature of ref.~\cite{De-83}
\begin{equation}
{\overline E}\equiv E({\bf p})=(p^2+m_N^2)^{1/2}
\label{XX2}
\end{equation}
yielding
\begin{equation}
{\overline E}-E=E_S+({\cal E}+{\overline E}-m_N)+T_{A-1}^0.
\label{XX3}
\end{equation}
In sec.~\ref{sect4}, \ref{sect5} and \ref{sect6} we shall return to make 
use of eq.~(\ref{XX3}).

As discussed in ref. \cite{DaMcDoSi-90}, it is then a simple matter 
to identify in the $({\cal E},p)$ plane the
domain that is compatible with the conservation of energy and momentum for 
the process whose cross section is given by eq.~(\ref{II.1}).
It turns out that
\begin{equation}
{\cal E}^+ \leq {\cal E} \leq {\cal E}^-,
\end {equation}
where 
\begin{equation}
{\cal E}^\pm = M^0_A + \omega - {\sqrt {p^2 + \MA}} \ -
\ {\sqrt {(p \pm q)^2 + m_N^2}}
\label{II.9}
\end {equation}
and $p = | {\bf p} |$, $q = |{\bf q}|.$
Clearly the larger (smaller) excitation energy
of the residual nucleus occurs when its momentum 
is parallel (antiparallel) to the momentum $\bf q$ transferred
to the system. Moreover $\E^-$ 
is bounded according to 
\begin{equation}
%controllare com math. c'e' il dubbio su un segno
0 \leq {\cal E}^-\leq \omega - E_S - \left [   
\sqrt {{ q}^2 + (M^0_{A-1}+m_N)^2} - (M^0_{A-1} + m_N )\right ],
\end {equation}
since it reaches its maximum for a momentum
\begin{equation}
p=p_{\rm max}=q\frac{M^0_{A-1}}{M^0_{A-1}+m_N}.
\label{II.9.1}
\end{equation}

\begin{figure}[ht]
\vskip10cm
\includegraphics{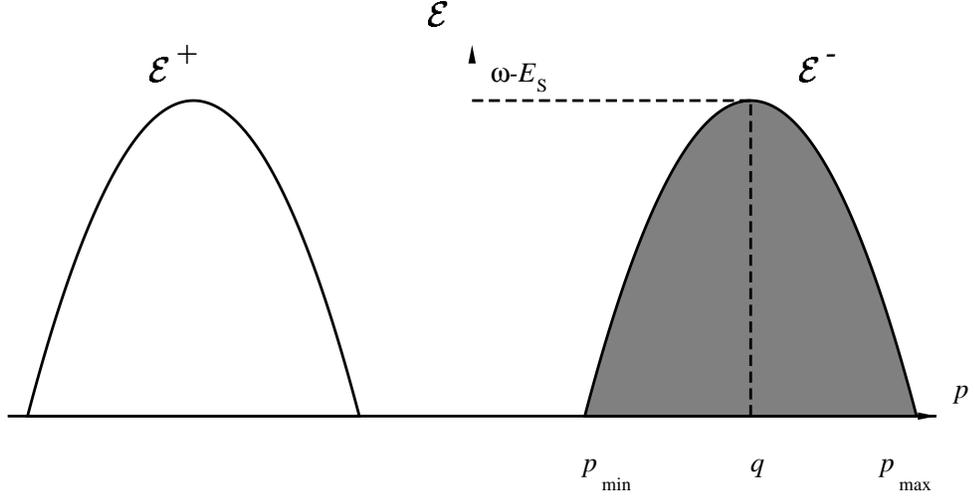}
\caption{\protect\label{Fig2} The curves $\E^-$ and
$\E^+$ given by eq.~(\protect\ref{II.9}) for a kinematics
corresponding to the left side of the QEP. A
semi-inclusive process can only occur within the shaded area.}
\end{figure}

\begin{figure}
\vskip10cm
\includegraphics{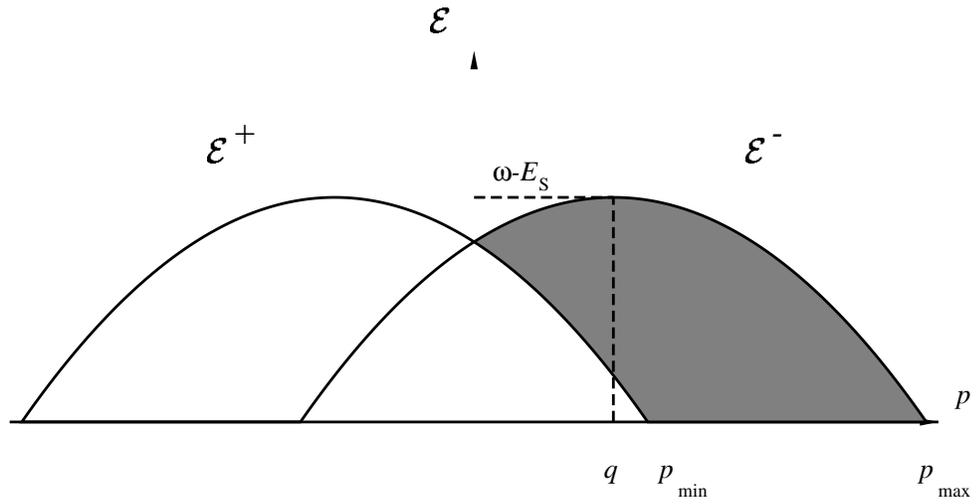}
\caption{\protect\label{Fig3} As in fig. \protect\ref{Fig2}, but for a
kinematics corresponding to the right side of the QEP.}
\end{figure}

To provide an appreciation for the boundaries given by eq.~(\ref{II.9}) 
these are displayed in fig.~\ref{Fig2} for $q$ and $\omega$ such that
${\cal E}^\pm{(0)} <0$,
namely for a point to the left of the quasielastic peak (QEP) of
the inclusive response: the shaded domain in
fig. \ref{Fig2}
is where the semi-inclusive process can occur.
On the other hand, for a point 
to the right of the RFG response peak, where 
${\cal E}^\pm (0) > 0\;,$
the allowed region is the one shown in 
fig. \ref{Fig3}.

For use later we now briefly revisit the scaling variable, defined 
as \cite{DaMcDoSi-90}
\begin {equation}
y = - p_{{\rm{min}}},
\label{II.14x}
\end {equation}
$p_{{\rm{min}}}$ being the smaller of the two real roots of the equation
\begin{equation}
{\cal E}^- (p) = 0.
\label{II.14}
\end {equation}
Thus the scaling variable is (up to a sign) the smallest momentum at zero 
missing energy that a particle can have inside the nucleus in order to be 
active in a semi-inclusive process. 
The larger solution of eq.~(\ref{II.14}), customarily denoted by $Y$,
will also be needed in the following.
A direct calculation then yields 
\begin{eqnarray}
\lefteqn{\left.{y(q, \omega)\atop Y(q, \omega)}\right\}={1
\over {2 W^2}}}
\label{II.15}\\
&\times\left\{ (M^0_A+\omega)\sqrt{W^2-(M^0_{A-1}+m_N)^2}
\sqrt{W^2-(M^0_{A-1}-m_N)^2} \right. \nonumber\\
&\left. \mp q (W^2+\MA -m_N^2) \right\},
\nonumber
\end{eqnarray}
where $W=\sqrt{(M^0_A + \omega)^2 - q^2}$.
Notice that $y$ and $Y$ are defined in the range
\begin {equation}
\omega_t \leq \omega \leq q,
\end{equation}
where
\begin{equation}
\omega_t = E_S + {\sqrt{q^2 + (M^0_{A-1} + m_N)^2}} - (M^0_{A-1} 
+ m_N) \ 
\label{omegat}
\end{equation}
is the threshold energy.
Also for
\begin{equation}
\omega = \omega_{\rm max}= E_S+{\sqrt{q^2 + m_N^2}} - m_N  
\end{equation}
the scaling variable vanishes.

\begin{figure}[ht]
\vskip10cm
\includegraphics{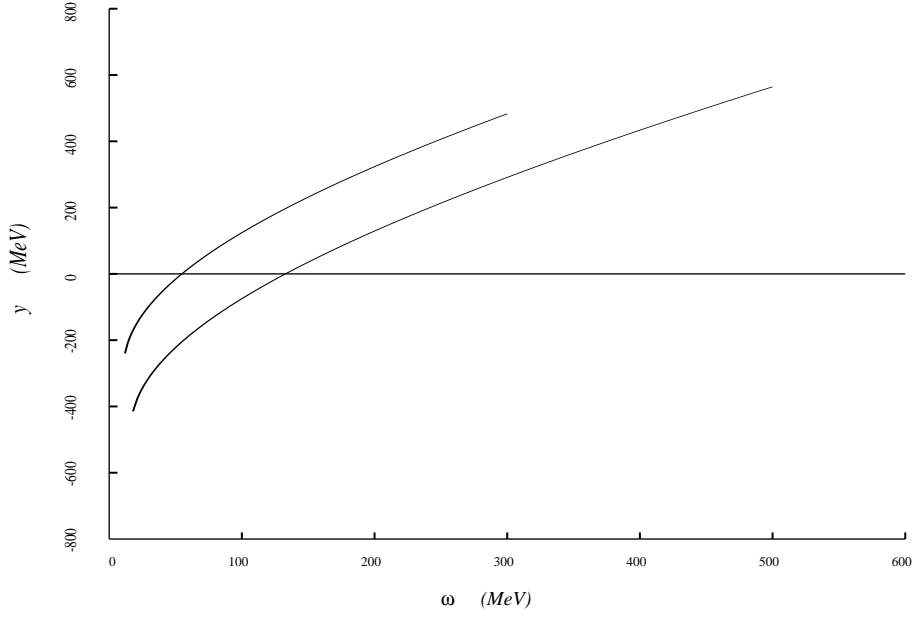}
\caption{\protect\label{Fig4}
Frequency behaviour of the scaling variable
for fixed $|\bf q|$. In the upper curve $|{\bf q}|$ = 300 MeV/c, while
in the lower $|{\bf q}|$ = 500 MeV/c (in the figure
$A=16$ and a separation energy of 8 MeV is assumed).}
\end{figure}
In fig. \ref{Fig4} we display the monotonic behaviour of $y$
versus
$\omega$ for fixed $q$. 
It reaches its minimum (negative)
value $y_{\rm min}=-p_{\rm max}$ (see eq.~(\ref{II.9.1})) for
$\omega=\omega_t$, while its maximum 
(positive) value
is reached on the light front at $y_{\rm max}=y(q,q)$.
Thus $y_{{\rm{min}}}$ can be arbitrarily decreased by simply 
increasing $q$. In contrast, $y_{\rm max}$
is bound
by the limiting value
\begin {equation}
\displaystyle\lim_{q \to \infty} y_{{\rm{max}}} (q) = {1 \over {2 M^0_A}}
\left\{ (M^0_A)^2 - \MA \right\}\sim m_N.
\end{equation}

\section{The impulse approximation \label{sect3}}

\begin{figure}[ht]
\vskip9cm
\includegraphics{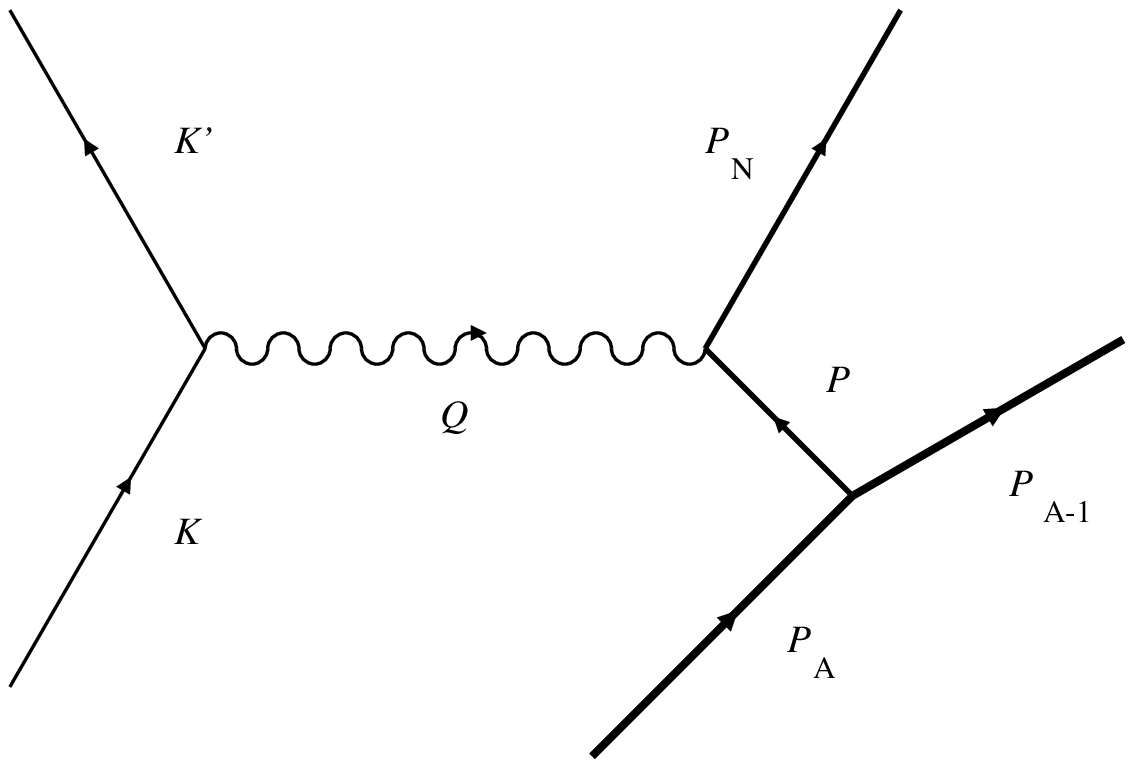}
\caption{\protect\label{Fig5} The PWIA interpretation of the box in fig.
\protect\ref{Fig1}.}
\end{figure}

In this section we further elaborate the cross section given in 
eq.~(\ref{II.1}). This we do within the context of the PWIA which offers the 
simplest interpretation of the box of fig. \ref{Fig1} through the diagram 
displayed in fig. \ref{Fig5} in which the virtual photon is absorbed by a 
single nucleon inside the nucleus, the rest of the system acting as a 
spectator. 
We first consider the hadronic tensor
\begin {equation}
W^{\mu\nu} = < A |\hbox{$ {\hat J^\dagger}$}^\mu | F > < F | \hat J^\nu | A >,
\label{Alfredo}
\end {equation}
$|F>$ being the continuum final state of the system system with $A$ nucleons.
In the impulse approximation only one-body hadronic currents are considered. 
These may be written
\begin{equation}
\hat J^\mu (Q) = \int {{d^3 p} \over {(2\pi)^3}}\intt{p^\prime}\ {{m_N} \over 
{E({\bf p})}}\ {{m_N} \over {E({\bf p}^\prime)}}\ 
< {\bf p}^\prime| J^\mu(Q) | {\bf p} > \hat 
a^\dagger_{p^\prime}
\hat a_{p},
\label{II.24a}
\end {equation}
where
\begin{equation}
< {\bf p}^\prime| J^\mu(Q) | {\bf p} >=(2\pi)^3\delta({\bf p}^\prime-{\bf
p}-{\bf q})\,({\bf p}+{\bf q}|\Gamma^\mu(Q)|\bf {p})
\label{II.24c}
\end{equation}
and $\Gamma^\mu (Q)$ is the, in general off-shell, electromagnetic $\gamma NN$ 
vertex operator. On-shell one has
\begin{equation}
( {\bf p} + {\bf q}| \Gamma^\mu | {\bf p}) = \overline u (P + Q) 
\left [ \gamma^\mu F_1 (Q^2) + {{i \sigma^{\mu\nu}} \over {2 m_N}}\ Q_\nu
F_2 (Q^2)\right] u(P),
\label{II.24b}
\end {equation}
as usual letting $F_1$ and $F_2$ be the Dirac and the Pauli form factors, 
whereas off-shell various prescriptions are commonly used \cite{De-83} one 
of which is discussed in the next section.

The basic ingredient of the PWIA is embodied in the structure of the 
final state, which is factorized according to
\begin{equation}
<F| \hat a^\dagger_{{\bf p}^\prime} = <{\bf p}_N|{\bf p}^\prime>
< B |=(2\pi)^3{E({\bf p}^\prime)\over m_N}\delta({\bf p}_N-{\bf p}^\prime)<B|,
\label{II.26}
\end {equation}
$| B>$ being any of the excited states (including the ground state) that
the residual $A - 1$ nucleus might be left in
(the factors in eq.~(\ref{II.26}) preserve the normalization of $<F|$ ).
In eq.~(\ref{II.26}) the plane-wave description of the outgoing nucleon 
ignores its past history on its way out of the
nucleus. We consider only kinematics where effects such as the distortion 
of the final-state nucleon wave function are expected to be small and thus, 
while our goal in the present work is focused on a model-to-model comparison, 
the results we obtain should be expected to be applicable at high $q$ for 
roughly quasielastic kinematics. 
With the help of eqs.~(\ref{II.24a}) and (\ref{II.26})
the hadronic tensor is found to be
\begin{equation}
W^{\mu\nu} = \left({{m_N} \over {E_{\bfm{\scriptstyle 
p}_N-\bfm{\scriptstyle q}}}} \right)^2< A |
\hat a^\dagger_{\bfm{\scriptstyle p}_N-\bfm{\scriptstyle q}} | B > < B | 
\hat a_{\bfm{\scriptstyle p}_N-\bfm{\scriptstyle q}} | A >
{\cal W}^{\mu\nu} ({\bf p}_N-{\bf q}, {\bf p}_N),
\label{II.29}
\end{equation}
\begin{equation}
{\cal W}^{\mu\nu} ({\bf p}_N - {\bf q},{ \bf  p}_N) = ( {\bf p}_N- {\bf q} | 
{\Gamma^\dagger}^\mu | 
{\bf p}_N 
)\,( {\bf p}_N | \Gamma^\nu | {\bf p}_N - {\bf q} )
\end {equation}
being the single-nucleon tensor.

We can now recast the semi-inclusive cross section in the following form
\begin{eqnarray}
\lefteqn{{{d^4 \sigma} \over {d \varepsilon^\prime d \Omega_e d E_N d 
\Omega_N}}
= {{4 \alpha^2} \over {(2 \pi)^3}} {{m_e^2 m_N p_N k^\prime} \over
\varepsilon} {1 \over {Q^4}} 
\left ( {{m_N} \over {E({\bf p}_N-{\bf q}) }} \right)^2 } 
\label{II.30}\\
&&\times \eta^{\mu\nu}{\cal W}_{\mu\nu} ({\bf p}_N
- {\bf q}, {\bf p}_N) \sum_{\beta}{{< A | \hat a^\dagger_{\bfm{\scriptstyle p}_N
-\bfm{\scriptstyle q}} | A-1,\beta> < A-1,\beta| 
\hat a_{\bfm{\scriptstyle p}_N-\bfm{\scriptstyle q}} | A >} \over {< 
A-1,\beta | A-1,\beta>}} 
\nonumber\\
&&\quad\quad  \times\delta (\omega + M^0_A - E_N - E_{A-1}^\beta),
\nonumber
\end{eqnarray}
where $\beta$ runs over all possible quantum numbers of the $A-1$ 
nucleus, or, by exploiting completeness,
\begin{eqnarray}
\lefteqn{{{d^4 \sigma} \over {d\varepsilon^\prime d\Omega_edE_Nd\Omega_N}}
= {{4\alpha^2} \over {(2\pi)^3}} {{m_e^2m_Np_Nk^\prime} \over \varepsilon}}
\label{II.32}\\
&&\times{1 \over {Q^4}} \eta^{\mu\nu} \left({{m_N}
\over {E({\bf p_N-q)} }}\right)^2 
{\cal W}_{\mu\nu} ({\bf p}_N-{\bf q}, {\bf p}_N) 
\nonumber\\
&&\times<A | \hat a^\dagger_{\bfm{\scriptstyle p}_N-\bfm{\scriptstyle q}} 
\delta (\omega + M^0_A - E_N -
E^0_{A-1}- \hat {\cal H}) 
\hat a_{\bfm{\scriptstyle p}_N-\bfm{\scriptstyle q}} | A >.
\nonumber
\end{eqnarray}

In the above the Hamiltonian $\hat {\cal H}=\hat H-E_{A-1}^0$, 
acting on the quantum states of the $A-1$ system, yields the excitation 
energies ${\cal E}$. Setting $E = E_N - \omega$, the energy of a nucleon 
inside the nucleus as above, and $\mu = M^0_A - E_{A-1}^0$, the chemical 
potential we introduce  the spectral function (our notation is chosen so as 
to maintain the standard definitions in the literature \cite{mahaux92})
\begin{equation}
\tilde  S ({\bf q}, q_0) = {1 \over (2\pi)^3} {m_N \over E({\bf q})}
{< A | \hat a^\dagger_q \delta [q_0 + \hat {\cal H} - \mu] \hat a_q
| A > \over < A | A >} ,
\label{II.33}
\end{equation}
which obeys the model-independent relations
\begin{equation}
\ssi\int \limits_{-\infty}^\infty d E\, \tilde  S ({\bf p}, E) = n(p)
\label{III.9}
\end {equation}
and
\begin{equation}
\ssi\int d^3p\int \limits_{-\infty}^\infty d E\,\tilde  S ({\bf p}, 
E) = A.
\label{III.10}
\end {equation}
In eq.~(\ref{III.9}) $n(p)$ is the momentum distribution 
of the nucleons in the nucleus.

Then it is an easy matter to re-express the cross section as 
\begin{equation}
{{d^4 \sigma} \over {d \varepsilon^\prime d \Omega_e dE_N d \Omega_N}}
= E_N p_N \left ({{d \sigma} \over {d \Omega}} \right)_{{\rm{eN}}}
\tilde  S ({\bf p}, E),
\label{II.36}
\end{equation}
where 
\begin{eqnarray}
\left ({d\sigma \over d\Omega} \right)_{\rm eN} &=&{4 \alpha^2
\over Q^4} m^2_e {k^\prime \over \varepsilon} {m^2_N \over E({\bf p})
E_N} \eta^{\mu\nu} {\cal W}_{\mu\nu} ({\bf p}, {\bf p} + {\bf q})
\nonumber\\
\label{II.35}
&=&\sigma_M\frac{1}{\cos^2\frac{\theta}{2}}
\frac{m_e^2}{\varepsilon\varepsilon^\prime}\eta^{\mu\nu} 
\frac{M^2_N}{E({\bf p})E_N}{\cal W}_{\mu\nu} ({\bf p}, {\bf p} + {\bf q})
\end{eqnarray}
is the cross section for elastic $eN$ scattering and
\begin{equation}
\sigma_M={4\alpha^2\over Q^4}(\varepsilon^\prime)^2\cos^2{\theta\over2}
\label{III.12.3}
\end{equation}
is usual the Mott cross section. Projections of the single-nucleon tensor
${\cal W}_{\mu\nu}$, both on- and off-shell, are discussed in the next 
section.

\section{The inclusive cross section, response functions and sum rule
\label{sect4}}

To recover the inclusive cross section in the quasielastic region from 
eq.~(\ref{II.36})
an integration should be performed over the outgoing nucleon's variables
$E_N$ and $\Omega_N$ and a sum should be made over protons and neutrons. 
In the  PWIA, however, it is natural to perform the 
integration in the $({\cal E},p)$ plane. Henceforth we shall write the 
arguments of the spectral function as $({\bf p},{\cal E})$ rather than 
$({\bf p},E)$ as in the last section. Clearly the variables are related by 
eq.~(\ref{XX1}). For the purpose of performing the integration we first add an 
integration over $p$, compensated by a $\delta$-function, to get
\begin{equation}
{{d^2 \sigma} \over {d\Omega_e d \varepsilon^\prime}} = 
\int d E_N \int d^3p \left( {{d\sigma}\over {d \Omega}}
\right)_{{\rm{eN}}}
\tilde  S ({\bf p},{\cal E}) \delta \left(E_N - {\sqrt{({\bf p + q})^2
+ m_N^2}} \right).
\label{III.12}
\end{equation}
Next we introduce  an azimuthal average over the single-nucleon cross 
section according to
\begin{equation}
< \sigma_0^{eN} > = {1 \over {2\pi}} \int \limits^{2\pi}_0 d \phi
\left({{d\sigma} \over {d\Omega}} ({\bf p, q})\right)_{{\rm{eN}}},
\label{65}
\end{equation}
exploit the $\delta$-function to integrate over $E_N$ and 
use the energy conservation to translate the remaining angular 
integration into one over $\E$.
We thus obtain
\begin{equation}
{d^2\sigma \over d \Omega_e d \varepsilon^\prime} = \int \limits_\Sigma
 d p  \,d \E\, p^2{E_N\over pq} 2\,\pi
< \sigma_0^{eN} > \tilde  S ({\bf p},{\cal E}),\label{III.13}
\end {equation}
where $\Sigma$ denotes the integration domain in the $(\E,p)$ plane
(shaded regions in figs. \ref{Fig2} and \ref{Fig3}), or, more explicitly,
\begin{equation}
\int \limits_\Sigma dp  d \E\dots=
\theta(-y)\int \limits_{-y}^Ydp\int \limits_0^{\E^-}d\E\dots
+\theta(y)\left[\int \limits_0^ydp\int \limits_{\E^+}^{\E^-}d\E\dots
+\int \limits_y^Ydp\int \limits_0^{\E^-}d\E\dots\right].
\label{III.13.1}
\end{equation}

Let us now split the inclusive cross section into its
longitudinal and transverse components setting
\begin{equation}
\frac{d^2\sigma}{d\Omega d\epsilon^\prime}=\sigma_M
\left\{v_L R_L(q,\omega)+v_T R_T(q,\omega)\right\},
\label{n1}
\end{equation}
with
\begin{eqnarray}
v_L&=&\left({Q^2\over q^2}\right)^2\\
v_T&=&\frac{1}{2}\left|{Q^2\over q^2}\right|+\tan^2\frac{\theta}{2}
\end{eqnarray}
and, in terms of the components of the hadronic tensor in eq.~(\ref{Alfredo}),
\begin{eqnarray}
R_L&=&W^{00}\\
R_T&=&W^{11}+W^{22}.
\end{eqnarray}

From eq.~(\ref{III.13}) it is then clear that the spectral function, which 
incorporates the nuclear dynamics, is channel independent and that 
the longitudinal or transverse nature of the response is reflected
only in the single-nucleon cross section. After some algebraic 
manipulations of eq.~(\ref{II.35}) on-shell the latter can be cast in the form
\begin{equation}
<\sigma_0^{eN}>=\sigma_{M}
\frac{m_N}{E({\bf p})}\frac{m_N}{E_N}\left[v_L{\cal R}_L+v_T{\cal R}_T\right],
\label{n3}
\end{equation}
with 
\begin{eqnarray}
{\cal R}_L&=&\frac{\kappa^2}{\tau}\left\{\chi^2\left[F_1^2(\tau)+\tau 
F_2^2(\tau)\right]+\left[F_1(\tau)-\tau F_2(\tau)\right]^2\right\}\label{Z1}\\
{\cal R}_T&=&\chi^2\left[F_1^2(\tau)+\tau F_2^2(\tau)\right]+2\tau
\left[F_1(\tau)+F_2(\tau)\right]^2.\label{Z2}
\end{eqnarray}
The factors in front of the right-hand side of eq.~(\ref{n3}) are in accord 
with standard
notation \cite{Cab93,De-83}. Note also that the dimensionless variables 
$\kappa=q/2 m_N$ and $\tau=-Q^2/4 m_N^2=(q^2-\omega^2)/4 m_N^2$ 
have been introduced (in the following we shall also use the dimensionless 
transferred energy $\lambda=\omega/2 m_N$). Moreover in the above 
\begin{equation}
\chi\equiv\frac{p}{m_N}\sin\theta,
\label{n4}
\end{equation}
$\theta$ being the angle between $\bf p$ and $\bf q$.
For convenience we also quote an equivalent expression for the 
longitudinal single-nucleon response, {\it viz.}
\begin{equation}
{\cal R}_L=\frac{\kappa^2}{\tau}\left\{G_E^2(\tau)+W_2(\tau)\chi^2\right\},
\label{nx}
\end{equation}
where 
\begin{equation}
W_2(\tau)=\frac{1}{1+\tau}\left[G_E^2(\tau)+\tau G_M^2(\tau)\right]\;
\label{n5}
\end{equation}
is given in terms of the familiar Sachs nucleon form factors $G_E$ and $G_M$.

All of the relations in eqs.~(\ref{Z1}, \ref{Z2}, \ref{nx}) hold for a 
process with the initial 
nucleon in the vacuum and hence on-shell. Since the nucleon inside the 
nucleus is in general off-mass-shell, a prescription is required for its 
current. Here we follow the one labeled CC1 by 
De Forest \cite{De-83} which sets
\begin{eqnarray}
\overline{\omega}&\equiv&E_N-\overline{E}\label{ZZ1}\\
\overline{\tau}&\equiv&\frac{q^2-\overline{\omega}^2}{4 m_N^2},\label{ZZ2}
\end{eqnarray}
where $\overline{E}$ is given by eq.~(\ref{XX2}), 
and yields for the single nucleon's responses the expressions 
\begin{eqnarray}
{\cal R}_L&=&\frac{\kappa^2}{\overline{\tau}}\left\{\chi^2
\left[F_1^2(\tau)+\overline{\tau }
F_2^2(\tau)\right]+\left[F_1(\tau)-\overline{\tau} F_2(\tau)\right]^2\right\}\\
{\cal R}_T&=&\chi^2\left[F_1^2(\tau)+\overline{\tau}
 F_2^2(\tau)\right]+2\overline{\tau}
\left[F_1(\tau)+F_2(\tau)\right]^2,
\end{eqnarray}
{\it i.e.} in this off-shell prescription the energy transfer $\omega$ 
is shifted in the single-nucleon 
responses, but not inside the form factors where it is kept unaltered.
The amount of the shift will be discussed further in sec.~\ref{sect6}. We
note in passing \cite{Cab93} that the CC2 prescription of 
De Forest \cite{De-83} 
yields the same results as above for ${\cal R}_L$ but not for ${\cal R}_T$. 
With the above definitions the nuclear responses finally read 
\begin{equation}
R_{L,T}(q,\omega)=\int \limits_\Sigma
 d p  \,d \E\, p^2{E_N\over pq} 2\,\pi 
\tilde  S ({\bf p},{\cal E})
\frac{m_N}{\overline{E}}
\frac{m_N}{E_N}
{\cal R}_{L,T}(q,{\overline{\omega}},p,\E).
\label{nr}
\end{equation}

Worth emphasizing here is the dependence of the single-nucleon responses 
upon $\chi$, in turn entailing a dependence upon the 
angle $\theta$ and, through momentum conservation,
upon $p$ and $\E$. As a consequence, as 
already stressed in \cite{Al-al-88} and at variance with the non-relativistic 
case, the single-nucleon responses cannot be moved 
outside the integral over the $\E$  and $p$ variables, {\it i.e.,} they
cannot be factorized.

With the responses fixed, a relativistic extension of 
the Coulomb sum rule and higher-order energy-weighted integrated 
responses can be obtained. In the non-relativistic case the Coulomb sum rule
$\Xi_0(q)$ is extracted from the data 
simply by dividing for $G_E^2$ and one obtains for large $q$ the well-known
result $\Xi_0(q)\to 1$. In the relativistic case, where 
the integration over $\omega$ is extended up to the light front 
and physical nucleons are considered, an appropriate reduction factor
must be introduced in order to fulfill the sum rule.

Actually, as we shall discuss in detail in the following sections, a function
$H_L(q,\omega)$ can be defined such that 
\begin{equation}
{{\cal R}_L(q,\omega,p,\theta)\over H_L(q,\omega)}\sim 1
\label{nh}
\end{equation}
in the whole domain where the spectral function is appreciably
different from zero. As a consequence a Coulomb sum rule can be set up
according to the prescription (see below)
\begin{equation}
\Xi_0(q)=\int\limits_{\omega_t}^{q}d\omega
\int_\Sigma dp\,d\E \tilde S(p,{\cal E}){{\cal 
R}_L(q,{\overline{\omega}},p,\theta)\over H_L(q,{\overline{\omega}})}\sim
 \int\limits_{\omega_t}^{q}d\omega\int_\Sigma 
dp\,d\E \tilde S(p,{\cal E}),
\label{nss}
\end{equation}
with $\omega_t$ given by eq.~(\ref{omegat}). 

The difficulty in establishing the above relation stems from the fact that
the integrals do not extend 
over the whole $(\E,p)$ plane -- in such a case one would
invoke the sum rule for the spectral function, namely eq.
(\ref{III.10}), to validate the Coulomb sum rule at once. However, 
when $q$ is sufficiently large, the integration domain actually encompasses
the whole region of the $(\E,p)$ plane where the nuclear spectral function is
appreciably different from zero.

To illustrate this point we define the quantities
\begin{eqnarray}
\E_{1,3}&=&M^0_A+\omega_t-\sqrt{\MA+p^2}-\sqrt{m_N^2+(q\pm p)^2}\label{x13}\\
\E_{2,4}&=&M^0_A+q-\sqrt{\MA+p^2}-\sqrt{m_N^2+(q\mp p)^2},\label{x24}
\end{eqnarray}
\begin{figure}
\vskip15.5cm
\includegraphics{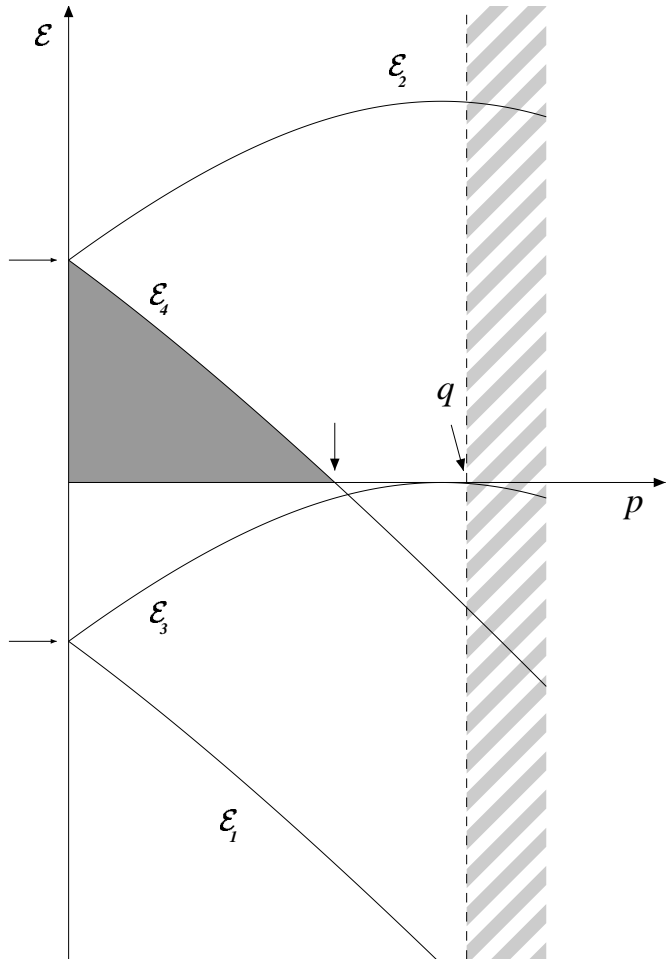}
\caption{\protect\label{Fig6}The boundaries of the integration region 
for the sum rule.}
\vskip0.5cm
\end{figure}
which are nothing but the boundaries $\E^\pm$ evaluated
for the minimum allowed value of $\omega$, namely $\omega_t$,
and for $\omega$ on the light cone, respectively. Obviously 
no contribution to the integral in eq.~(\ref{nss}) arises from the regions
$\E<\E_1$ and $\E>\E_2$. On the other hand, denoting the extreme values of 
$\cos\theta$ by $\cos\theta_{\rm min}$ and $\cos\theta_{\rm max}$, so that 
$\cos\theta_{\rm min}\leq\cos\theta\leq\cos\theta_{\rm max}$, then one can 
show that for $\E_3\leq\E\leq\E_4$ the integration range will have 
$\cos\theta_{\rm min}=-1$ and $\cos\theta_{\rm max}=+1$. Under these 
circumstances the full sum rule can be attained; however, for $\E_1\leq\E<\E_3$ 
and $\E_4<\E\leq\E_2$ the full range of integration will not be available 
and one should expect the sum rule to be under-saturated. A final constraint 
is of course that $\E\geq 0$. 

The typical behaviour of eqs.~(\ref{x13}) and (\ref{x24}) is displayed in 
fig. \ref{Fig6}. Several points on these curves are of particular significance 
and are indicated in the figure with arrows. Two include the ${\cal E}$-values 
at $p=0$: in particular,
\begin{equation}
\E_2(0)=\E_4(0)=M^0_A+q-M^0_{A-1}-\sqrt{m_N^2+q^2}\stackrel{q\to\infty}
{\longrightarrow}m_N-E_S,
\label{n11}
\end{equation}
which lies well above the scale of nuclear energies normally considered.
Another is the value of $p$ where $\E_4(p)=0$, which occurs for
\begin{eqnarray}
p&=&\overline{p}=-\frac{q\left((M^0_A)^2+\MA+2M^0_Aq-m_N^2\right)}{
2M^0_A\left(M^0_A+2q\right)}
\label{n12}\\
&+&\frac{M^0_A+q}{2M^0_A\left(M^0_A+2q\right)}\sqrt{\left[
(M^0_A)^2-\MA -m_N^2+2M^0_Aq\right]^2-4\MA m_N^2}.
\nonumber
\end{eqnarray}
For large momentum transfers the above equation becomes
\begin{equation}
\overline{p}\stackrel{q\to\infty}{\longrightarrow}\frac{(M^0_A)^2-\MA}{
2M^0_A}\sim m_N,
\label{n13}
\end{equation}
again well beyond the scale of momenta characterizing the nuclear momentum 
distribution. 

In view of the magnitude of the above-quoted limiting values, it follows that 
for sufficiently large $q$ the spectral function is expected to be concentrated
in the shaded region of fig.~\ref{Fig6}. Thus, if a convenient form for the 
normalization factor $H_L$ is found, then the sum  rule over the spectral
function automatically generates a sum rule over the longitudinal response. 
Within the context of the PWIA violation of the sum rule might then imply 
that significant contributions in the spectral function occur at very 
high missing energy and/or at very high missing momentum (both, of course, 
are absent in the models considered in the present work). 

It is important to realize that the factor $H_L$, that accounts
both for the kinematical 
factors of the single-nucleon cross section (solely due to relativity) and for
the off-shellness of the single-nucleon response,
turns out to be largely model-independent.
Its on-shell form will be given in the next section, while 
its off-shell extension will be discussed in sec.~\ref{sect6}.

In concluding this section we also define the reduced longitudinal response 
\begin{equation}
r_L(q,\omega)\equiv \frac{R_L(q,\omega)}{H_L(q,\omega)}
\label{redresp}
\end{equation}
and the dimensionless energy-weighted moments according to
\begin{equation}
\Xi_n(q)=\int_0^q d\omega \lambda^n r_L(q,\omega),
\label{mom}
\end{equation}
where, as before, $\lambda=\omega/2 m_N$.
We shall examine some of these moments in the next two sections, together 
with the variance
\begin{equation}
\sigma(q)=\sqrt{\Xi_2(q)-\left(\Xi_1(q)\right)^2}.
\label{var}
\end{equation}

\section{The Free Relativistic Fermi Gas\label{sect5}}

The free RFG should be considered as a paradigm for all fermion many-body
systems. In its ground state the RFG has all its single-particle levels 
occupied by on-mass-shell nucleons up to the Fermi momentum $k_F$, a 
configuration with zero total momentum. Note that all states considered here 
in constructing this model are completely covariant, something that is 
usually given up in considering models with confinement, as in the next 
section. We write for the energy of the RFG initial state
\begin {equation}
M_A^{0,{\rm{RFG}}} = 4\sum_{k \leq k_F} {\sqrt{{\bf k}^2 + m_N^2}}.
\end {equation}
Here $A$ is a very large even integer to become infinite in the end and 
we consider only $N=Z$ nuclei in the present work. 
If a particle is removed from the surface of the RFG then the system thus
obtained, which is clearly in its ground state (indicated by ``0''), will 
have an energy 
\begin{eqnarray}
E^{0,{\rm RFG}}_{A-1} &=& M_A^{0,{\rm{RFG}}} - {\sqrt{k_F^2 + m_N^2}}\\
&=&{\sqrt{ \left(M^{0,{\rm RFG}}_{A-1}\right)^2 + k_F^2}}\\
&\equiv& T_{A-1}^{0,{\rm RFG}} + M_{A-1}^{0,{\rm RFG}}.
\label{III.2a}
\end {eqnarray}
It then follows that that the daughter nucleus ground-state mass is given by
\begin{equation}
M^{0,{\rm RFG}}_{A-1} = {\sqrt{(M_A^{0,{\rm RFG}})^2 - 2 M_A^{0,{\rm RFG}}
{\sqrt{k_F^2 + m_N^2}} + m_N^2}}.
\end {equation}
Hence the separation energy of the RFG may be written
\begin{eqnarray}
E_S^{{\rm{RFG}}} &=& M_{A-1}^{0,{\rm RFG}} + m_N - M_A^{0,{\rm RFG}} \nonumber\\
&=& \sqrt{(M_A^{0,{\rm RFG}})^2 - 2 M_A^{0,{\rm RFG}} {\sqrt{k_F^2 + m_N^2}} 
+ m_N^2} + m_N - M_A^{0,{\rm RFG}}\nonumber\\
&=& -T_F-T_{A-1}^{0,{\rm RFG}}\label{xxx}\\
&\stackrel{A\to\infty}{\longrightarrow}& -T_F,\nonumber
\end {eqnarray}
where $T_F\equiv (k_F^2+m_N^2)^{1/2}-m_N\cong -k_F^2/2m_N$. Note that 
$E_S^{\rm RFG}$ is negative, in contrast to the physical separation energy 
which must be positive.

We may also consider the $(A-1)$ Fermi gas in an excited state (namely with a
hole inside the Fermi sphere) with
an energy
\begin{eqnarray}
{\cal E}^{{\rm{RFG}}} &= &[E_{A-1} - E^0_{A-1}]^{\rm RFG} \nonumber\\
&=& {\sqrt{k_F^2 + m_N^2}} - {\sqrt{p^2 + m_N^2}},
\label{III.5}
\end{eqnarray}
which, since $0 \leq p\leq k_F$, is non-negative as it should be. Note 
also that this result together with eq.~(\ref{XX3}) imply that 
$[{\overline E}-E]^{\rm RFG}=0$, that is, as expected the struck nucleon 
is on-shell in the RFG model.

\begin{figure}
\vskip10.5cm
\includegraphics{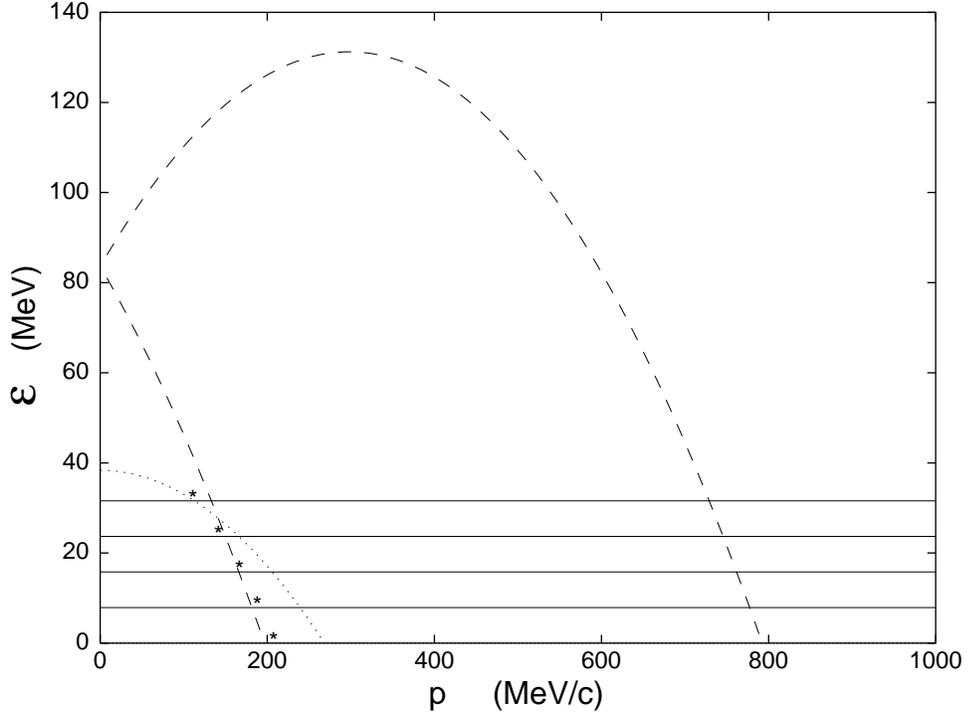}
\caption{\protect\label{Fig7}
The support of the spectral function of a shell model nucleus
with $A=140$ and $N_{\rm max}=4$ (parallel horizontal
straight lines) with
$\omega_0=7.9$ MeV. Also shown is the support
of the spectral function of the RFG as given by eq.~(\protect\ref{III.5})
with $\k=1.112$ fm$^{-1}$ corresponding
to the fictitious nucleus $^{140}$Yb (see eq.~(\protect\ref{IV.20})).
The stars correspond to the rms value of the momentum of a particle in
a shell of given $N$ (see eq.~(\protect\ref{IV.18})).
Finally the allowed domain for the exclusive process is shown for
$q=300$ MeV and $\omega=100$ MeV (to the right of the response peak).}
\end{figure}

The curve obtained from eq.~(\ref{III.5}), and displayed in fig. \ref{Fig7} 
together with the region kinematically accessible
to a semi-inclusive process for typical values of $q$ and
$\omega$, defines the support of the RFG spectral function ${\tilde 
S}^{\rm RFG}({\bf p},{\cal E})$, {\it i.e.} the one-dimensional domain in the 
$({\cal E}, p)$ plane where it is non-vanishing.

As a consequence the RFG inclusive cross section is expected to reach its 
maximum, but for the influence of the single-nucleon physics, 
for $\omega$ and $q$ such that the domain of integration in the 
$({\cal E}, p)$ plane includes the whole of ${\tilde  S}^{\rm RFG}({\bf p},
{\cal E})$. For this to occur it is required that at $p$=0  
\begin{equation}
{\cal E}^- (0) = {\cal E}^+ (0) = {\cal E}^{\rm{RFG}}(0),
\end{equation}
which requires that
\begin{equation}
M^0_A + \omega - M^0_{A-1} - {\sqrt{q^2 +  m_N^2}} 
= {\sqrt{k^2_F + m_N^2}} - m_N.
\end{equation} 
Hence the well-known expression
\begin{equation}
\omega = \omega_{\rm QEP}\equiv{\sqrt{q^2 + m_N^2}} - m_N=|Q^2|/2m_N
\label{omegqep}
\end{equation}
follows for the maximum of the RFG response.

To calculate ${\tilde 
S}^{\rm RFG}({\bf p},{\cal E})$ it helps to recall the Lehmann 
representation of the 
relativistic single-particle Green's function for an infinite, homogeneous 
system, which, disregarding antinucleonic components, reads
\begin{eqnarray}
\lefteqn{G_{\alpha\beta}(p,E({\bf p}))= \frac{1}{(2\pi )^3}
\left({{\pslash + m_N} \over 
{2Ep}} \right)_{\alpha\beta}{{m_N} \over {E({\bf p})}}}
\label{III.6}\\
&&\times\left\{< A| \hat a_{p s} {1 \over {E({\bf p}) - \hat {\cal H} - \mu
+ i \eta}} \hat a^\dagger_{p s} | A> 
\right.
\nonumber\\
&&\left.
+ < A | \hat a^\dagger_{p s} {1 \over {E({\bf p})+ \hat {\cal H} - \mu
- i \eta}} \hat a_{p s} | A>\right\}.
\nonumber
\end{eqnarray}
Then from eq.~(\ref{II.33}) the following 
connection between the hole part of the fermion propagator and the 
spectral function
\begin{equation}
\Im\, G^h_{\alpha\beta} (p) = \pi \left({\pslash + m_N\over 2E({\bf p})}
\right)_{\alpha\beta}
\tilde  S ({\bf p}, {\cal E})
\label{III.7}
\end{equation}
is found to hold true. For the RFG, whose states are normalized according to
\begin{equation}
<{\bf p}|{\bf q}>=\Omega{E({\bf p})\over m_N}\delta_{p,q},
\end{equation}
$\Omega$ being the volume enclosing the system, rather than by 
eq.~(\ref{II.1.1.1}), one gets from eq.~(\ref{III.7}):
\begin{eqnarray}
{\tilde  S}^{\rm RFG} ({\bf p}, {\cal E}) &=& \frac{\Omega}{(2\pi)^3} 
\theta (k_F - p)\delta (E({\bf p})-\sqrt{p^2+m_N^2})\nonumber\\
&=&\frac{\Omega}{(2\pi)^3} \theta (k_F - p)\delta(\E-\sqrt{\k^2+m_N^2}
+\sqrt{p^2+m_N^2}),
\label{III.8}
\end{eqnarray}
which is non-vanishing in the $(\E,p)$ plane only along the curve 
defined by eq.~(\ref{III.5}).

To get the covariant RFG inclusive response we only need to insert 
eq.~(\ref{III.8}) into eq.~(\ref{III.12}), obtaining
\begin{eqnarray}
\lefteqn{{d^2\sigma\over d\Omega_e d\varepsilon^\prime}\Biggm|_{\rm RFG}=
\Omega\intt{p}\left({d\sigma\over d\Omega}\right)_{\rm eN}}
\label{III.12.1}\\
&&\theta(\k-p)\delta\left(
\omega-\left[\sqrt{({\bf p}+{\bf q})^2+m_N^2}-\sqrt{p^2+m_N^2}\right]\right),
\nonumber
\end{eqnarray}
which is indeed recognized as the Pauli-unblocked RFG cross section.
Note that the connection between exclusive and inclusive
processes for the RFG  can be established, in the PWIA framework, only in 
the Pauli unblocked region. 

We now briefly consider the $y$-scaling properties of the  RFG. In this model 
the onset of scaling is not induced by the behaviour of $Y$, which becomes 
larger than any characteristic scale in the problem for large $q$, but by
the spectral function, which fixes the upper integration limit over $p$
to be $\k$. Note furthermore that for the RFG the lower integration limit 
is not $y$, but rather the point where the support of the RFG spectral 
function intersects the boundary $\E^\pm$. Defining
\begin{equation}
y_{\rm RFG}\equiv m_N \left[ \lambda\sqrt{1+\frac{1}{\tau}} -\kappa \right] ,
\label{aux11}
\end{equation}
we find that this intersection point occurs at $p=|y_{\rm RFG}|$. The sign 
may be understood by making use of the $\psi$-scaling variable introduced 
in ref. \cite{Al-al-88}:
\begin{equation}
\psi=\frac{1}{\sqrt{\xi_F}}[2\theta(\lambda-\lambda_0)-1]
\left\{ \sqrt{(1+\lambda)^2+\frac{1}{\tau}(\tau-\lambda)^2}-(1+\lambda)
 \right\}^{1/2},
\label{psi}
\end{equation}
where 
\begin{equation}
\xi_F=T_F/m_N=\sqrt{1+\eta_F^2}-1\cong \eta_F^2/2
\end{equation}
since $\eta_F\equiv k_F/m_N\cong 1/4$ is small. Here
\begin{equation}
\lambda_0=\frac{\omega_{\rm QEP}}{2m_N} 
=\frac{1}{2}\left[\sqrt{1+4\kappa^2}-1\right]
=\tau_0=\frac{-Q_0^2}{4m_N^2}>0.
\label{ugh3}
\end{equation}
For the RFG one finds that $-1<\psi<1$ and $\psi=0$ 
at the quasielastic peak whose position is given by $\lambda_0$ 
(see eq.~(\ref{omegqep})). The inverse of eq.~(\ref{psi}) is given by
\begin{eqnarray}
\lambda &=& \frac{1}{2}\left[\sqrt{1+\left\{2\kappa+\psi\sqrt{
\xi_F(2+\xi_F\psi^2)}\right\}^2}-(1+\xi_F\psi^2)\right] \label{ugh1}\\
&=& \lambda_0+\left( {\frac{2\kappa}{\sqrt{1+4\kappa^2}}} \right) 
  {\frac{1}{2}}\eta_F\psi+{\cal O}[\eta_F^2].
\label{ugh1x}
\end{eqnarray}
The latter also allows us to obtain
\begin{equation}
\tau = \lambda_0 \left[ 1-\left( {\frac{2\kappa}{\sqrt{1+4\kappa^2}}} \right) 
  \eta_F\psi+{\cal O}[\eta_F^2] \right]
\label{taueqn}
\end{equation}
and these expressions permit us to rewrite eq.~(\ref{aux11}) in the forms
\begin{eqnarray}
y_{\rm RFG} &=& \sqrt{2\xi_F}\psi\left\{ 1+\frac{1}{2} \xi_F\psi^2 
  \right\}^{1/2} \label{aux11xy}\\
&=& \frac{m_N}{\kappa} \left\{ \lambda \left[ 
  1+\xi_F\psi^2 \right] -\tau \right\} \label{aux11x}\\
&=& m_N \sqrt{2\xi_F}\psi +{\cal O}[\eta_F^2]. \label{aux11xx}
\end{eqnarray}
The first of these three forms shows that if the reduced response 
scales with $\psi$ then it also scales with $y_{\rm RFG}$. Furthermore, from 
the first and second equations here it is clear that $y_{\rm RFG}=0$ at the 
QEP where $\lambda=\lambda_0=\tau=\tau_0$ and $\psi=0$. Indeed, as the 
third form shows most clearly, to the extent that we are dealing with 
dilute systems ($\eta_F^2<<1$) we have that $y_{\rm RFG}\cong m_N \eta_F 
\psi$.

For comparison, in the $A\rightarrow\infty$ limit the solution to 
eqs.~(\ref{II.14x}--\ref{II.14}) yields 
\begin{eqnarray}
y &=& 2m_N \left[ \sqrt{\lambda_1 (1+\lambda_1)} - \kappa \right]\\
&=& \frac{m_N}{\kappa (1+\frac{y}{2q})}\left[ \lambda_1 
-\tau_1 \right],
\label{yyyy}
\end{eqnarray}
where $\lambda_1\equiv \lambda-E_S/2m_N$ and $\tau_1\equiv 
\kappa^2-{\lambda_1}^2$. Later (see eq.~(\ref{XY2})) we shall introduce the 
dimensionless energy $\lambda^\prime$ and thus for completeness we note 
also that $\lambda_1=\lambda^\prime+\xi_F/2$. For large values of $q$ at 
fixed $y$ (the scaling limit) it is clear that the term $y/2q$ in 
eq.~(\ref{yyyy}) may be neglected and that $y$ has roughly the same form as 
$y_{\rm RFG}$, except that the peak position is shifted by $E_S$, as 
discussed in ref.~\cite{DaMcDoSi-90}.

For very light nuclei (very small $k_F$) the distinction between $y$ and 
$y_{\rm RFG}$ is not very important. Indeed, for extreme cases such as 
$^3$He the most important contributions to the inclusive cross section 
when not too far from the QE peak arise from a limited region at small $p$ 
and small ${\cal E}$. Hence emphasizing the extreme (and this is used in 
the definition of $y$; see eqs.~(\ref{II.14x}, \ref{II.14})) is reasonable 
in that case.  However, for heavy nuclei important contributions arise from 
other regions in the $(p,{\cal E})$--plane. As fig.~\ref{Fig7} shows, the 
RFG concentrates all of its strength along the dotted curve in the figure 
and therefore the intersection of the kinematic boundary with this curve 
provides the correct definition (for this model) of what we have called 
$y_{\rm RFG}$ in eq.~(\ref{aux11}). As we shall see in the next section, 
models such as the HM distribute the strength still differently and 
consequently may require definitions of yet other scaling variables. In the 
present work we limit our attention to $y$ and $y_{\rm RFG}$, the latter being 
well-approximated up to multiplicative normalizing constants by $\psi$, 
as well as to a new scaling variable denoted $\psi^\prime$ to be discussed 
in the next section.

We complete our review of the RFG by quoting the expression for the inclusive
cross section \cite{Al-al-88}
\begin{equation}
\frac{d^2\sigma}{d\Omega d\epsilon^\prime}=\frac{{\cal N}\sigma_M}{
4m_N\kappa}S(\psi)\left\{v_L U_L+v_T U_T\right\},
\label{rfgrl}
\end{equation}
where
\begin{equation}
U_L=\frac{\kappa^2}{\tau}\left(G_E^2+W_2\Delta\right)
\label{ul}
\end{equation}
and 
\begin{equation}
U_T=2\tau G_M^2(\tau)+W_2(\tau),
\label{ut}
\end{equation}
with
\begin{eqnarray}
\lefteqn{\Delta=\frac{\tau}{\kappa^2}\Bigl\{
\frac{1}{3}\left[\epsilon^2_F+\epsilon_F(1+\xi_F\psi^2)+(1+\xi_F
\psi^2)^2\right]}
\label{IV.74}\\
&&+\lambda\left[\epsilon_F+(1+\xi_F\psi^2)\right]+\lambda^2\Bigr\}-(1+\tau).
\nonumber
\end{eqnarray}
As usual for the inclusive cross section one should add copies of 
eq.~(\ref{rfgrl}) with ${\cal N}=Z$ and proton form factors to copies with 
${\cal N}=N$ and neutron form factors (see ref.~\cite{Al-al-88}). We shall use 
generic quantities ${\cal N}$ and $U_{L,T}$, although the proton$+$neutron 
sums should always be understood.

Worth noting is the analogy between $U_L$ defined above 
and ${\cal R}_L$ as given by eq.~(\ref{nx}). Their structure is similar: indeed
while in ${\cal R}_L$ there appears the factor $\chi$,
which implies a dependence upon $p$ and $\E$, in $U_L$ such a 
factor is replaced by $\Delta$, which is constant with respect to these 
variables. Thus $U_L$ may be viewed as a suitable average of ${\cal 
R}_L$ and is a natural ingredient to be used in setting up the reduction
factor $H_L$, which can only depend on $q$ and $\omega$ ($\kappa$ and 
$\lambda$). One finds that $\Delta$, which does not depend on $p$ and $\E$, 
in fact depends not only upon $\kappa$ and $\lambda$, but
upon the Fermi momentum $k_F$ as well: it thus still embodies some of the
dynamics of the nuclear medium.

The problem of setting up a reduced longitudinal response satisfying
the sum rule was solved for the RFG in ref.~\cite{Barbaro94} with the 
introduction of the following (see eq.~(\ref{redresp})): 
\begin{equation}
r_L^{\rm RFG} (q,\omega)=
\frac{ R_L^{\rm RFG} (q,\omega)}{H_L^{\rm RFG} (q,\omega)}
\label{sl}
\end{equation}
with
\begin{equation}
H_L^{\rm RFG} (q,\omega)=\frac{{\cal N}U_L}{J_L},
\label{sl1}
\end{equation}
where
\begin{eqnarray}
J_L &\equiv& {\frac{\kappa\eta_F^3}{2\xi_F}}
  {\frac{\partial\psi}{\partial\lambda}} =
  \left( {\frac{\eta_F}{\sqrt{2\xi_F}}} \right)^3 \sqrt{1+\frac{1}{2} 
  \xi_F\psi^2} \left( {\frac{\kappa^2}{\tau}} \right) \left[ 
  {\frac{1+2\lambda+\xi_F\psi^2}{1+\lambda+\xi_F\psi^2}} \right] \label{der1}\\
&=& \left( {\frac{\kappa^2}{\tau}} \right) \left[ 
  {\frac{1+2\lambda}{1+\lambda}} \right]+{\cal O}[\eta_F^2]. \label{der2}
\end{eqnarray}
As usual here, the generic form ${\cal N}U_L$ is used whereas the 
actual result must have $Z U_{Lp} + N U_{Ln}$. A little algebra then yields
\begin{equation}
r_L^{\rm RFG} (q,\omega)=\frac{3}{8m_N}(1-\psi^2)
\theta(1-\psi^2)\frac{\partial\psi}{\partial\lambda},
\label{hl}
\end{equation}
whose moments 
\begin{equation}
\Xi_n^{\rm RFG} (q)=\int\limits_{0}^{\infty} d\omega 
r_L^{\rm RFG} (q,\omega) \lambda^n
\label{ugh}
\end{equation}
are then easily evaluated using eq.~(\ref{ugh1}), which may be simplified 
when $\kappa>>\xi_F$ to the expression in eq.~(\ref{ugh1x}). One then 
immediately gets for the moments 
\begin{eqnarray}
\Xi_0^{\rm RFG}&=&1\qquad\qquad\qquad({\rm sum~rule})\\
\Xi_1^{\rm RFG}&\simeq&\lambda_0 \qquad\quad\qquad({\rm mean~energy})\\
\Xi_2^{\rm RFG}&\simeq&\lambda_0^2+\frac{1}{5}\frac{\kappa^2\eta_F^2}
{\left(1+4\kappa^2\right)}.
\end{eqnarray}
Finally, the variance of the RFG is then obtained according to
\begin{equation}
\sigma^{\rm RFG}=\sqrt{\Xi_2^{\rm RFG}-\left(\Xi_1^{\rm RFG}\right)^2}\simeq
\frac{1}{2\sqrt{5}}\frac{1}{m_N}\frac{q\k}{\sqrt{q^2+m_N^2}}
=\frac{1}{\sqrt{5}}\frac{\kappa\eta_F}{\sqrt{1+4\kappa^2}}.
\label{var_rfg}
\end{equation}

Having revisited the basics of the RFG and established in the RFG framework
the connection, for $q>2\k$, between inclusive and exclusive scattering, 
in the next section we shall set up a more realistic spectral function that 
in particular accounts for the confinement of the nucleons in the initial state.

\section{The hybrid model \label{sect6}}

In this section we derive a spectral function that accounts for
the confinement of the nucleons inside the nucleus in its initial state 
employing a simple, tractable model where the $A\to\infty$ limit may be 
relatively easily performed. The 
initial state is viewed as an assembly of independent particles moving 
non-relativistically in a harmonic oscillator well, namely
\begin{equation}
H_{HO}=\frac{p^2}{2m_N}+\frac{1}{2}m_N\omega_0^2 r^2-(N_{\rm max}+3/2)
\omega_0-E_S.
\label{how}
\end{equation}
The constant terms in the above set the energy scale in such a way that the
energy required to remove a nucleon from
the outermost shell just corresponds to the separation energy $E_S$. 
In the following $N = 2 n + \ell$ is the principal quantum number of an occupied
shell and $N_{\rm{max}}$ the highest value of $N$.
The energies of the excited state of the daughter nucleus are
\begin{equation}
M_{A-1} = M^0_A - m_N - \left( N - N_{\rm{max}}\right)
\omega_0+E_S
\end{equation}
and the excitation energy of the residual nucleus, measured with respect to the
ground state and in a frame where it is moving with momentum $- {\bf p}$,
are
\begin{eqnarray}
\E^{\rm HM}&=& {\sqrt{p^2 + (M_{A-1})^2}} - {\sqrt{p^2+\MA}}\nonumber\\
&\cong &{{p^2} \over {2 M_{A-1}}} + M_{A-1} - {{p^2} \over {2M^0_{A-1}}}
- M^0_{A-1} \nonumber\\
&\cong &(N_{{\rm{max}}} - N)\omega_0,
\label{IV.3}
\end {eqnarray}
where the recoil energy of the daughter nucleus has been neglected in the 
last expression.

In the same approximation the spectral function reads
\begin{equation}
{\tilde  S}^{\rm HM}({\bf p},{\cal E})=\sum_{N=0}^{N_{\rm max}}\delta\left[
\E-\left(N_{\rm max}-N\right)\omega_0\right]n_N(p),
\label{sfho}
\end{equation}
 where $n_N(p)$ is the momentum distribution of the $N^{\rm th}$ shell of 
the considered nucleus, {\it i.e.}
\begin{equation}
n_N(p) = \sum_{{n,\ell\atop (2n+\ell=N)}} | \varphi_{n\ell} (p)|^2 
\sum^\ell_{m=-\ell}|
Y_{\ell,m} (\hat{\bf p})|^2 = \sum_{{n,\ell\atop (2n+\ell=N)}}
{{2\ell+1}\over{4\pi}} | 
\varphi_{n\ell} (p)|^2,
\end{equation}
the radial harmonic oscillator wave functions in momentum space being
related
to the corresponding quantities in coordinates space $R_{n\ell}$ by
$(\nu = m_N \omega_0)$
\begin{equation}
\varphi_{n\ell} (p) = {\sqrt{{1 \over{\nu^3}}}} R_{n\ell} \left({p \over \nu}
\right).
\end{equation}
\begin{figure}
\vskip8cm\hskip1cm
\includegraphics{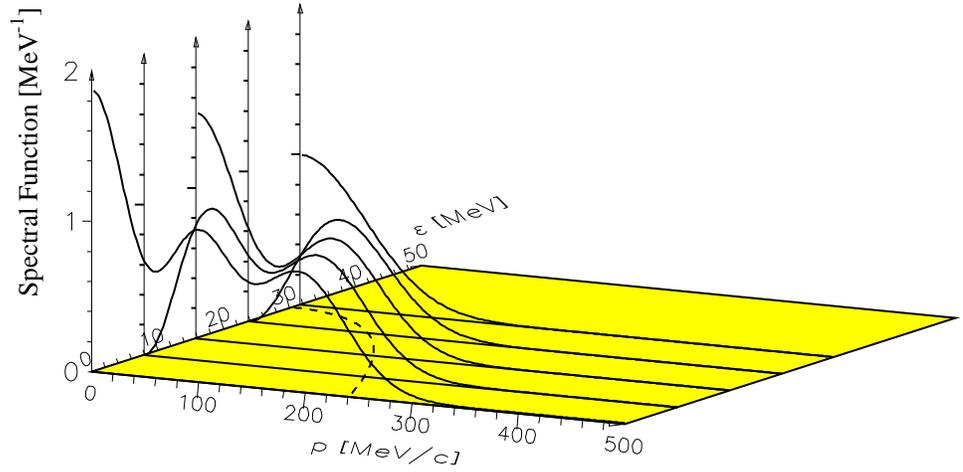}
\caption{\protect\label{Fig8}
The spectral function in MeV$^{-1}$ using harmonic oscillator wave 
functions for the ($N=Z$) nucleus
$A=140$ ($N_{\rm max}=4$) with $\omega_0=7.9$ MeV. Also displayed is the
support of the $\protect{\tilde  S}^{\rm RFG}$ for $\k=1.112$ fm$^{-1}$. The 
latter
confines the former, but for a small leakage, in a corner of the
$(\E,p)$ plane. Also shown is the kinematically allowed domain for the
exclusive process for $q=300$ MeV/c and $\omega=25$ MeV.
}
\end{figure}
Analytic expressions may be obtained for these quantities, although for 
the sake of brevity these are not given here. The support in the $(\E,p)$ 
plane of the harmonic oscillator spectral function 
${\tilde  S}^{\rm HM}({\bf p},E)$ is a set of lines, one for each shell $N$ of 
the daughter nucleus, all parallel to the $p$-axis. It is displayed in 
figs.~\ref{Fig7} and \ref{Fig8} for $N_{\rm max}=4$, which would correspond to 
the nucleus ${}^{140}$Yb, together with the region kinematically accessible to 
a semi-inclusive process for a choice of values of $\omega$ and $q$ yielding
a positive $y$. Also shown in the same figures is the corresponding 
curve from eq.~(\ref{III.5}) for the RFG for $\k=1.112$ fm$^{-1}$.

The two supports of the spectral functions displayed in figs.~\ref{Fig7} 
and \ref{Fig8} are indeed very different: 
the dramatic difference between $\tilde  S^{{\rm{RFG}}}$ and
$\tilde  S^{{\rm{HM}}}$ is fully apparent and it should not be surprising 
that the RFG and HM models yield drastically different predictions for 
exclusive processes.
One would like to fix $\k$ in such a way that the points where 
the support of the RFG spectral function crosses that of the harmonic 
oscillator HM coincide with the root-mean-square value of the momentum $p$
in the corresponding shell. For the harmonic oscillator 
the root-mean-square (r.m.s.) momentum in a given shell reads
\begin {equation}
{\sqrt{<p^2>_N}} = {\sqrt{m_N \omega_0 \left(N + {3\over 2}
\right)}}.
\end {equation}
Therefore eq.~(\ref{IV.3}) can be recast as follows
\begin {equation}
\E^{\rm HM} = \left(N_{{\rm{max}}} - N
\right) \omega_0 = \left(N_{{\rm{max}}} + {3\over 2}\right)
\omega_0 - {{<p^2>_N} \over {m_N}}
\label{IV.16}
\end {equation}
and then compared with the Fermi gas expression in the non-relativistic 
limit (see eq.~(\ref{III.5}))
\begin{equation}
\E^{\rm RFG} = {{k_F^2} \over {2m_N}} - {{p^2} \over {2m_N}},
\label{IV.17}
\end{equation}
viewing $<p^2>_N$ as a continuous variable. 
Owing to the factor ${1 \over 2}$ in front of 
the term $p^2/m_N$ in eq.~(\ref{IV.17})
one immediately realizes that no choice
of $\omega_0$ and $k_F$ can reconcile eqs.~(\ref{IV.16}) and (\ref{IV.17}).

For a chosen $\omega_0$ it is possible, however, by equating 
eqs.~(\ref{IV.16}) and (\ref{IV.17}) to determine a Fermi momentum $k_F^N$ 
such that the r.m.s. momentum in a given shell of the harmonic oscillator 
is identical to the Fermi gas value. One obtains
\begin{equation}
k_F^N = {\sqrt{m_N \omega_0}} {\sqrt{2N_{{\rm{max}}} - N + {3\over 2}}},
\label{IV.18}
\end {equation}
which implies that the inner (outer) shells of a 
nucleus are related to a Fermi gas with a larger (lower) density, 
respectively.

Quantitatively we illustrate this relation by displaying in table 
\ref{table1} the
values of $k_F^N$ for all the shells of six different
nuclei ($\omega_0 = 41/A^{1/3}$ MeV), where in our ``toy model" only 
$N=Z$ nuclei are considered.

In heavy nuclei the $k_F$ associated with the innermost shell $(N=0)$
are indeed close to the nuclear matter value $k_F^{nm} = 1.36$ 
fm$^{-1}$ whereas the outer shells relate to values of $k_F$ which
are about 30\% lower than $k_F^{nm}$; hence the correspondence between
$k_F$ and $\nu$, or better, between the density and the size of a nucleus,
can only be established on the average.
For example one might define for each nucleus
an average Fermi momentum according to
\begin{equation}
\overline{k_F} = {\sqrt{{{\sum_{N=0}^{N_{{\rm{max}}}} (N+1)(N+2)(k_F^N)^2}
\over{\sum_{N=0}^{N_{{\rm{max}}}} (N+1)(N+2)}}}},
\label{IV.20}
\end{equation}
$k_F^N$ being given by eq.~(\ref{IV.18}).

\begin{table}
%\centerline{
\begin{tabular}{||c||c||c||c||c||c||c||}
\hline
\hline
N&$^{16}$O&$^{40}$Ca&$^{80}$Zr&$^{140}$Yb&$^{224}?$&$^{336}?$\cr
\hline
\hline
           0
 &1.172 
 &1.261 
 &1.312 
 &1.345 
 &1.368 
 &1.385 
 \cr\hline
           1
 &0.990 
 &1.140 
 &1.221 
 &1.272 
 &1.307 
 &1.333 
 \cr\hline
           2
 &~
 &1.006 
 &1.123 
 &1.195 
 &1.243 
 &1.279 
 \cr\hline
           3
 &~
 &~
 &1.016 
 &1.112 
 &1.176 
 &1.222 
 \cr\hline
           4
 &~
 &~
 &~
 &1.023 
 &1.105 
 &1.162 
 \cr\hline
           5
 &~
 &~
 &~
 &~
 &1.029 
 &1.099 
 \cr\hline
           6
 &~
 &~
 &~
 &~
 &~
 &1.033 
 \cr\hline
\hline
 $\overline k_F$ 
 &1.039 
 &1.075 
 &1.097 
 &1.112 
 &1.123 
 &1.131 
\cr\hline
\hline
\hline
 $\alpha_F$
 &1.18
 &1.12
 &1.10
 &1.09
 &1.09
 &1.09
\cr\hline
\hline
\end{tabular}
%}
\vskip0.4cm
\caption{\protect\label{table1}
The Fermi momenta (all in fm$^{-1}$) associated with each shell $N$ 
of the harmonic oscillator. In the next-to-last row the Fermi momentum 
averaged according to eq.~(\protect\ref{IV.20}) is given. The final row 
gives the quantity $\alpha_F$ defined in eq.~(\protect\ref{alphaf}). }
\end{table}

\begin{figure}
\vskip9.5cm
\includegraphics{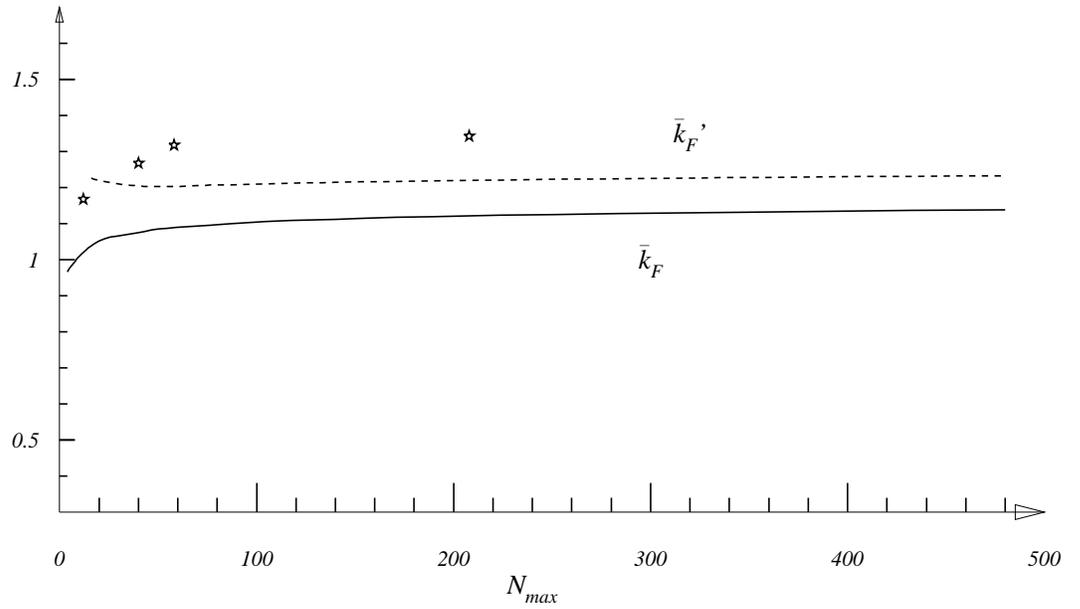}
\caption{\protect\label{Fig9}Values of  ${\overline k_F}$ and 
 ${\overline k_F}^\prime$ (in fm$^{-1}$) as defined by 
eqs.~(\protect\ref{IV.20}) and (\protect\ref{kfbarp}), respectively, versus 
$N_{\rm max}$. A few experimental values for nuclei with the same mass 
numbers are also shown. }
\end{figure}

The values of the Fermi momentum thus obtained are reported in table 
\ref{table1} and
displayed as a function of the mass number $A$ in fig. \ref{Fig9}
 where it is seen that the nuclear matter value 
of $1.36$ fm$^{-1}$ will just be reached for the innermost shells of very 
heavy nuclei. Note also that when available from the width of the 
quasi-elastic peak as measured in inclusive inelastic electron scattering 
the experimental values of $k_F$ appear to be typically about 20\%
higher than those predicted by eq.~(\ref{IV.20}). The curve labelled 
${\overline k_F}^\prime$ is discussed later.

Let us now compare the spectral function of the HM with that of the RFG,
which should be viewed as a $\delta$-function with its base lying on the
support curve of $\tilde S^{{\rm{RFG}}}$.
They are displayed in fig. \ref{Fig8} for the nucleus $A=140$ and for
$k_F=1.112$ fm$^{-1}$.

Once the spectral function is given -- see eq. (\ref{sfho}) --
we obtain the longitudinal response from eq.~(\ref{III.13}),
\begin{eqnarray}
\lefteqn{R_L^{\rm HM}(q,\omega)=}
\label{lrho}\\
&&{2\pi \over q} \sum_{N=0}^{N_{{\rm{max}}}}
\int\limits_\Sigma dp\,d\E\,p\,n_N(p) \frac{m_N^2}{E({\bf p})}
{\cal R}_L(q,\omega,p,\E)
\delta \left[ \E- \left(N_{{\rm{max}}} -
N\right) \omega_0\right],
\nonumber
\end{eqnarray}
which should be divided by an appropriate reduction factor as in 
eq.~(\ref{mom}) to obtain the sum rules. In this connection
the problem of the off-shellness of the single-nucleon response still has to be
faced. In fact only for the RFG has a reduction factor, namely the $H_L$ in
eq.~(\ref{sl1}), been obtained. 

\begin{figure}
\vskip17.6cm
\includegraphics{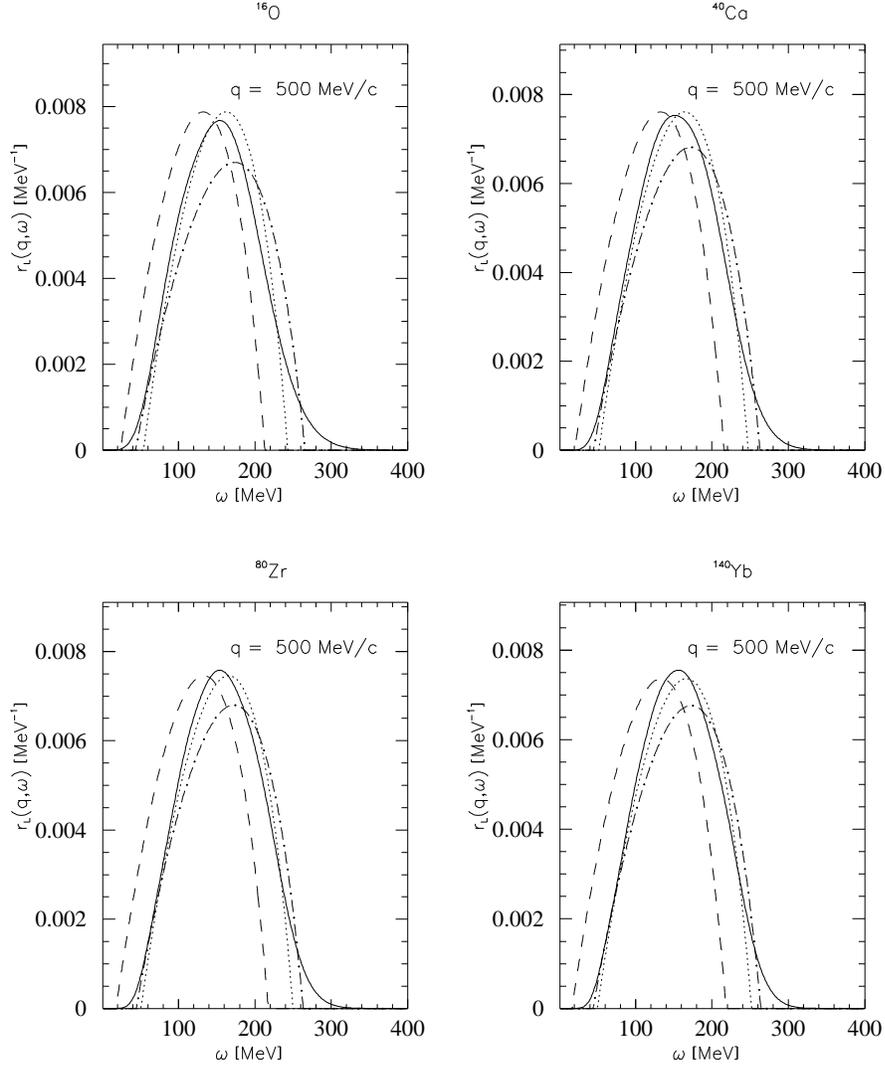}
\vskip-3cm
\caption{\protect\label{Fig10}Longitudinal reduced responses for different 
nuclei at $q=500$ MeV/c. 
Dashed lines: RFG, with $\k$ evaluated according to eq.~(\protect\ref{IV.20});
dotted lines: RFG, with $\k$ evaluated according to 
eq.~(\protect\ref{IV.20}) and the energy shift suggested in 
eq.~(\protect\ref{XY2}); dash-dotted lines: RFG, with the energy shift in 
eq.~(\protect\ref{XY2}) as well as with the scaled Fermi momentum given 
by eq.~(\protect\ref{kfbarp}); solid lines: hybrid model.
}
\end{figure}

We begin by noting that in general ${\cal R}_L$ depends upon the variables 
$p$ and $\E$ because the initial nucleon is not at rest and is off-shell, 
although in a low-density regime (as for 
nuclear 
matter) such a dependence is expected to be mild. 
Moreover, since for a finite nucleus the momentum distribution 
is also concentrated in a rather restricted region, in fact mostly within 
the Fermi sphere,
and the tail contribution is quite small, it might appear to be justified to 
assume the same reduction factor of the RFG, namely $H_L^{\rm RFG}(q,\omega)$, 
for the hybrid model as well.

\begin{figure}
\vskip17.6cm
\includegraphics{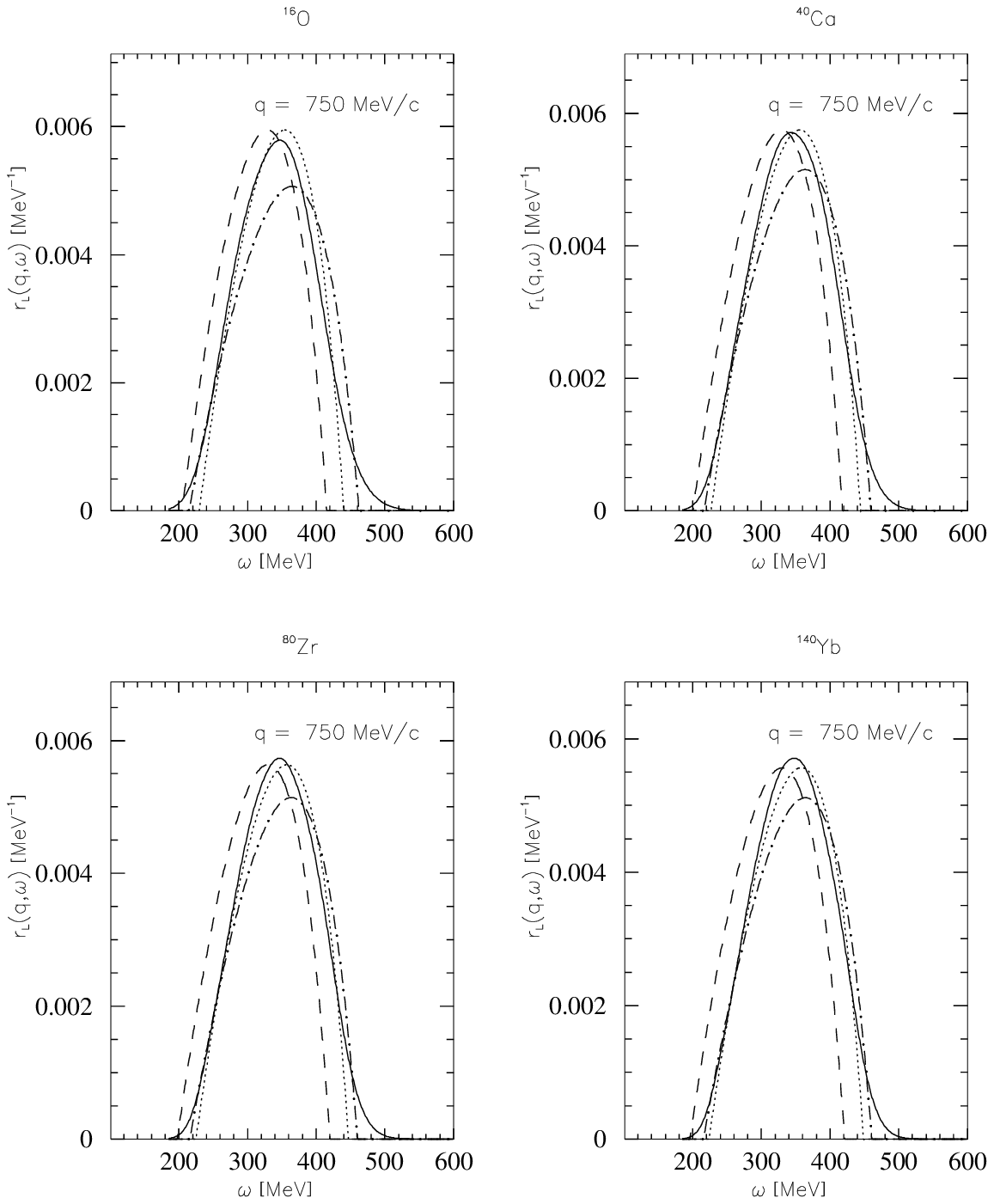}
\vskip-3cm
\caption{\protect\label{Fig11} Same as fig. \protect\ref{Fig10}, but for
$q=750$ MeV/c}
\end{figure}

A delicate point relates however to the energy scales 
of the RFG and of the hybrid model. 

Both are set by the separation 
energy $E_S$, but as discussed above in the RFG case the latter
is negative ($-T_F$, where $T_F = {\sqrt{m_N^2 + k_F^2}} -m_N$ is the 
kinetic energy of a particle
at the Fermi surface) while for the harmonic oscillator the separation
energy has been chosen to be positive, as it is for real nuclei. This 
forces the struck nucleon off-shell according to eq.~(\ref{XX3}) --- recall 
that $E$ is the correct ({\it i.e.,} off-shell) energy of the struck 
nucleon, whereas ${\overline E}$ in eq.~(\ref{XX2}) is the corresponding 
on-shell quantity for the same three-momentum $p$. 

\begin{figure}
\vskip17.6cm
\includegraphics{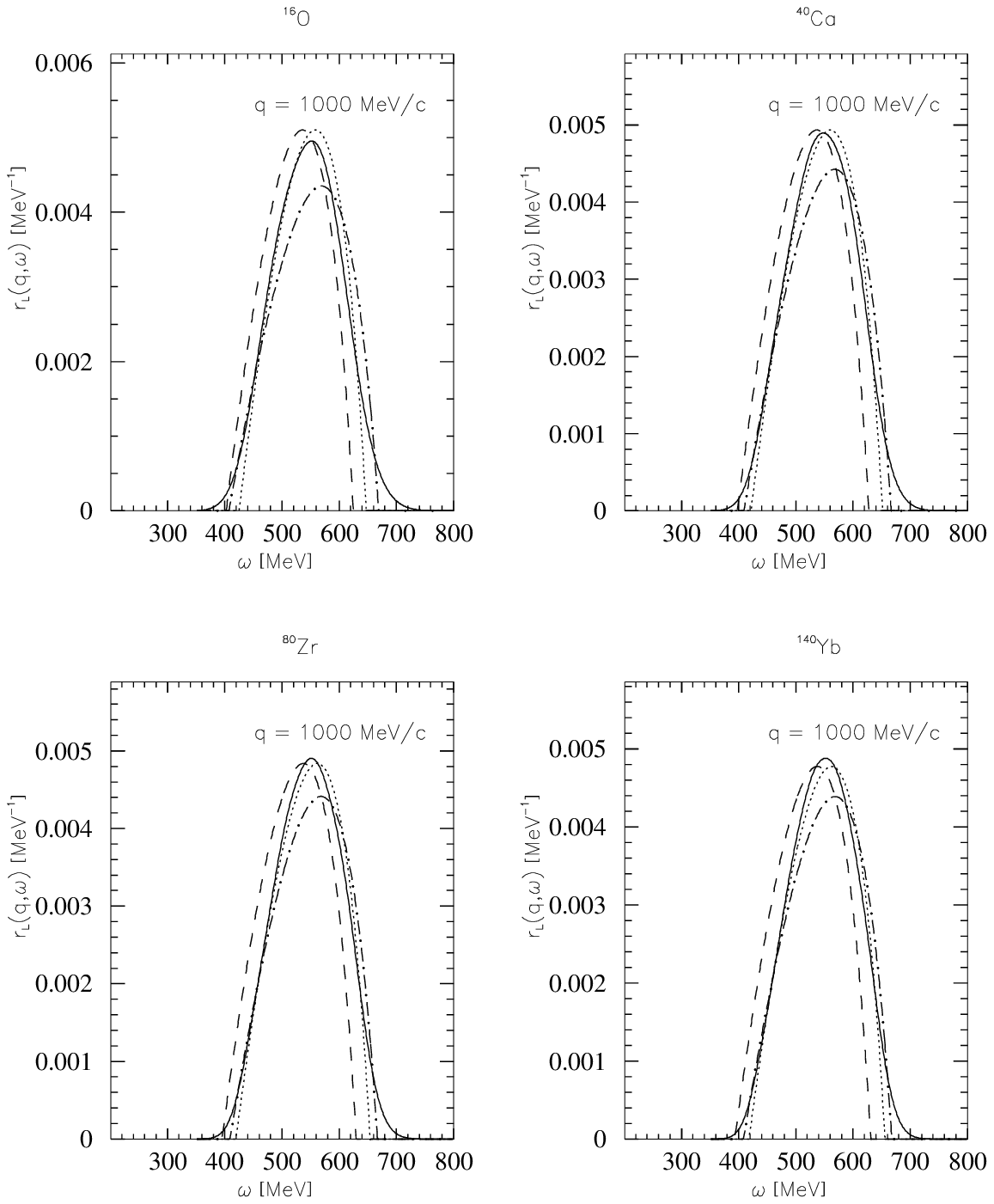}
\vskip-3cm
\caption{\protect\label{Fig12} Same as fig. \protect\ref{Fig10}, but for
$q=1000$ MeV/c}
\end{figure}

In our treatment using 
the HM we make two assumptions. First, we ignore the daughter nucleus 
recoil energy, $T_{A-1}^0\cong p^2/2M_{A-1}^0$, in eq.~(\ref{XX3}). Since
typically $p$ is less than $k_F$ where the spectral function plays a 
significant role 
and since our focus is on medium-weight nuclei (in fact, one of our goals 
has been to develop inter-model comparisons in the limit where $A\rightarrow 
\infty$), dropping this energy amounts to an error of less than an MeV 
under typical circumstances. Of course, for very light nuclei or for 
very large missing momenta one should be more careful with this 
contribution.

Second, in determining the reduction factor $H_L^{\rm HM}(q,\omega)$, which 
should only depend on $q$ and $\omega$, we use instead of the full variation in 
${\cal E}$ only the ``average value'' provided by the RFG as given in 
eq.~(\ref{III.5}). We then have
\begin{eqnarray}
{\cal E}+{\overline E}-m_N &\rightarrow& 
  {\cal E}^{\rm RFG}+{\overline E}-m_N \nonumber\\
&=& T_F = -E_S^{\rm RFG} \label{XY1}
\end{eqnarray}
and then eq.~(\ref{XX3}) yields
\begin{equation}
{\overline E}-E\rightarrow E_S+T_F = E_S-E_S^{\rm RFG}
\end{equation}
or, using eq.~(\ref{ZZ1}),
\begin{equation}
{\overline \omega}\rightarrow \omega^\prime\equiv\omega -[E_S-E_S^{\rm RFG}], 
\label{XY2}
\end{equation}
and $\lambda^\prime\equiv\omega^\prime/2m_N$; also ${\overline \tau}$ is then 
given by eq.~(\ref{ZZ2}) so that ${\overline \tau}\rightarrow 
\tau^\prime\equiv\kappa^2-{\lambda^\prime}^2$. Furthermore, in analogy with 
eq.~(\ref{psi}), we also define a ``shifted'' scaling variable:
\begin{equation}
\psi^\prime\equiv\frac{1}{\sqrt{\xi_F}}[2\theta(\lambda^\prime-
 \lambda_0)-1]\left\{ \sqrt{(1+\lambda^\prime)^2
 +\frac{1}{\tau^\prime}(\tau^\prime-\lambda^\prime)^2}-(1+\lambda^\prime)
 \right\}^{1/2}.
\label{psiprime}
\end{equation}
We note in passing that 
it appears \cite{Will96} that quasielastic data taken at intermediate 
energies in fact scale somewhat better when $r_F(q,\omega)$ is plotted 
as a function of $\psi^\prime$ than when plotted versus either $\psi$ or 
$y$. 
The new reduction
factor, denoted $H_L(q,\omega;E_S)$, is then obtained from an extension of 
eq.~(\ref{sl1}) to account for the off-shell nature of the EM vertex:
\begin{equation}
H_L (q,\omega;E_S)=\frac{{\cal N}U_L^{\rm off}}{J_L^{\rm off}},
\label{sl1off}
\end{equation}
where $J_L^{\rm off}$ is given by eqs.~(\ref{der1},\ref{der2}) with the above 
replacements of $\lambda\rightarrow\lambda^\prime$ and $\tau\rightarrow
\tau^\prime$ together with the CC1/CC2 form for the longitudinal 
projection of the off-shell EM current,
\begin{eqnarray}
U_L^{\rm off} &=& {\frac{\kappa^2}{\tau}} \left[ \left\{ G_E^2 
  +{\frac{1}{\tau^\prime}}(\tau-\tau^\prime)[ F_1^2-\tau\tau^\prime F_2^2 ]
  \right\} \right. \nonumber\\
  && \qquad \left. +\left\{ W_2 
  +{\frac{1}{\tau^\prime}}(\tau-\tau^\prime) F_1^2 
  \right\} \Delta (\lambda\rightarrow\lambda^\prime,\ \tau\rightarrow
  \tau^\prime) \right]
\label{uloff}
\end{eqnarray}
with $\Delta$ given by eq.~(\ref{IV.74}) after replacement to primed 
variables. As usual ${\cal N}U_L^{\rm off}$ should be taken to mean 
$Z U_{Lp}^{\rm off}+N U_{Ln}^{\rm off}$. Note that $H_L (q,\omega;E_S)$ 
reverts to the on-shell answer in eq.~(\ref{sl1}) when $E_S=E_S^{\rm RFG}$. 

Having fixed the definition of $H_L (q,\omega;E_S)$ 
the reduced response and the sum rules can be evaluated numerically. We 
display the HM reduced charge response in figs. \ref{Fig10}-\ref{Fig12} for 
$q = 500$, 750 and 1 GeV respectively (solid curves). In the figures the 
$E_S$ of the HM is set to 8 MeV 
and the $k_F$ of the RFG is fixed according to eq.~(\ref{IV.20}).

The basic features of the HM reduced response (as compared to that of the RFG
shown as dashed curves) emerging from the present analysis are:

\begin{description}
\item {i)} an upward shift in energy,
\item {ii)} ``tails'' at high and low $\omega$, and
\item {iii)} a somewhat larger half-width (see below).
\end{description}
\noindent All of these traits were of course to be expected and are easily 
understood: indeed
it is harder to pull nucleons out of a system when they are bound,
hence the shift in energy, and confined
nucleons have distributions extending to higher momenta because of the
uncertainty principle, hence the larger width and tails of the HM charge 
response. Notably these features appear to be largely unaffected by the 
momentum transferred to the nuclear system at least in the range up to 1 GeV.

\begin{figure}
\vskip17.4cm\hskip-1.2cm
\includegraphics{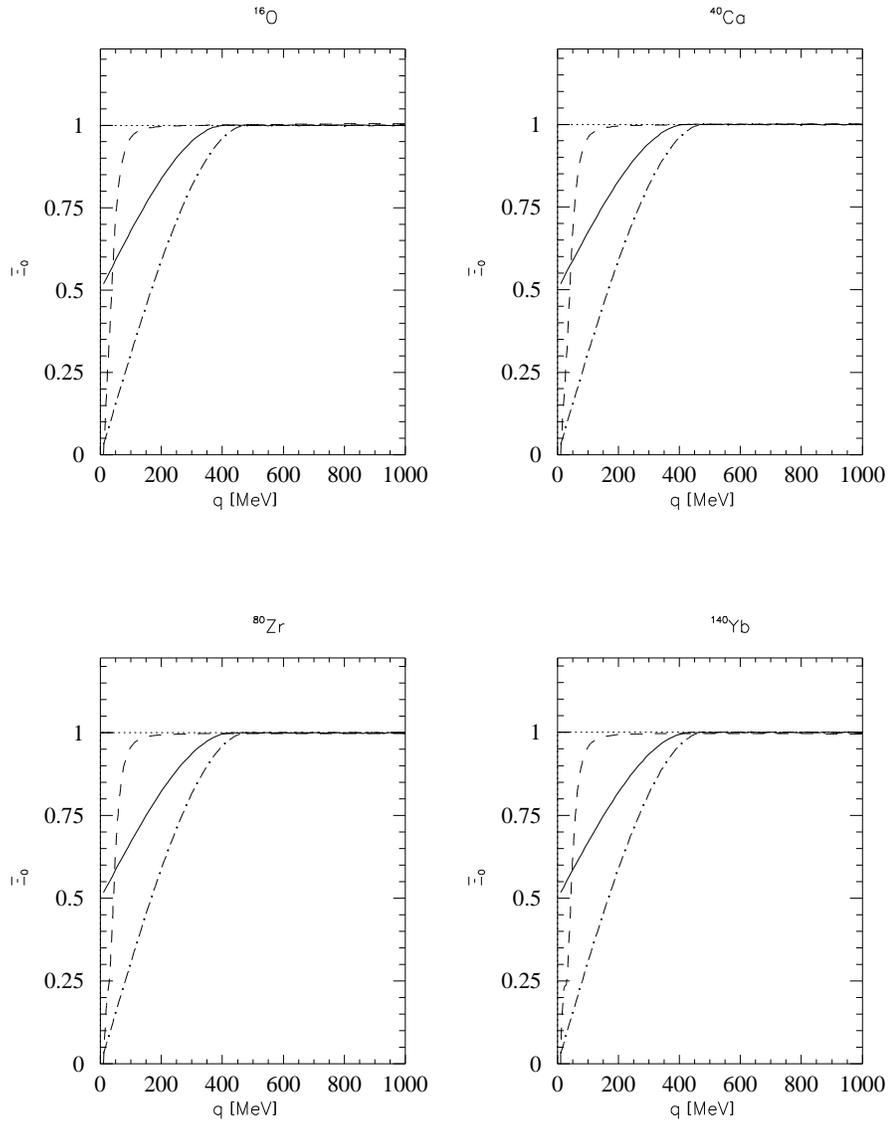}
\vskip-3cm\hskip3.2cm
\caption{\protect\label{Fig13}The normalized Coulomb Sum Rule.
Solid lines: non-Pauli-blocked RFG with integration from $0$ to $q$; 
dotted lines: non-Pauli-blocked RFG with integration from $E_S$ to $q$; 
dash-dotted lines: RFG with integration from $0$ to $q$ and with the 
inclusion of the Pauli principle; dashed lines: hybrid model.}
\end{figure}

\begin{figure}
\vskip17.4cm\hskip-1.2cm
\includegraphics{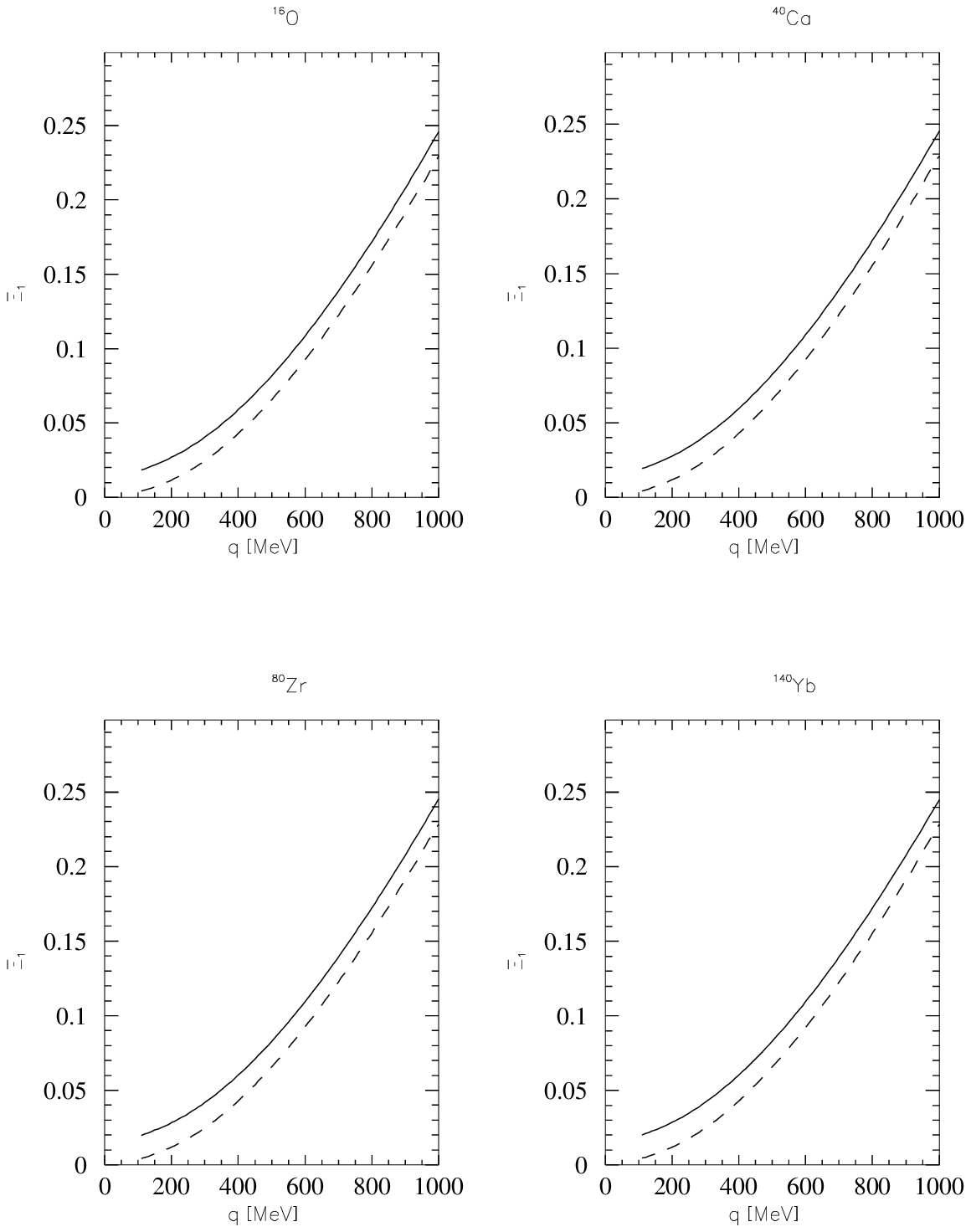}
\vskip-3cm\hskip3.2cm
\caption{\protect\label{Fig14}
The normalized energy-weighted sum rule.
Dashed lines: RFG, with $\k$ evaluated according to eq.~(\protect\ref{IV.20});
solid lines: RFG with $\k$ evaluated according to 
eq.~(\protect\ref{IV.20}) together with the energy shift suggested in 
eq.~(\protect\ref{XY2}) or RFG with the energy shift in 
eq.~(\protect\ref{XY2}) as well as with the scaled Fermi momentum given 
by eq.~(\protect\ref{kfbarp}) or hybrid model (the last three yield a 
single result).
}
\end{figure}

\begin{figure}
\vskip17.4cm\hskip-1.2cm
\includegraphics{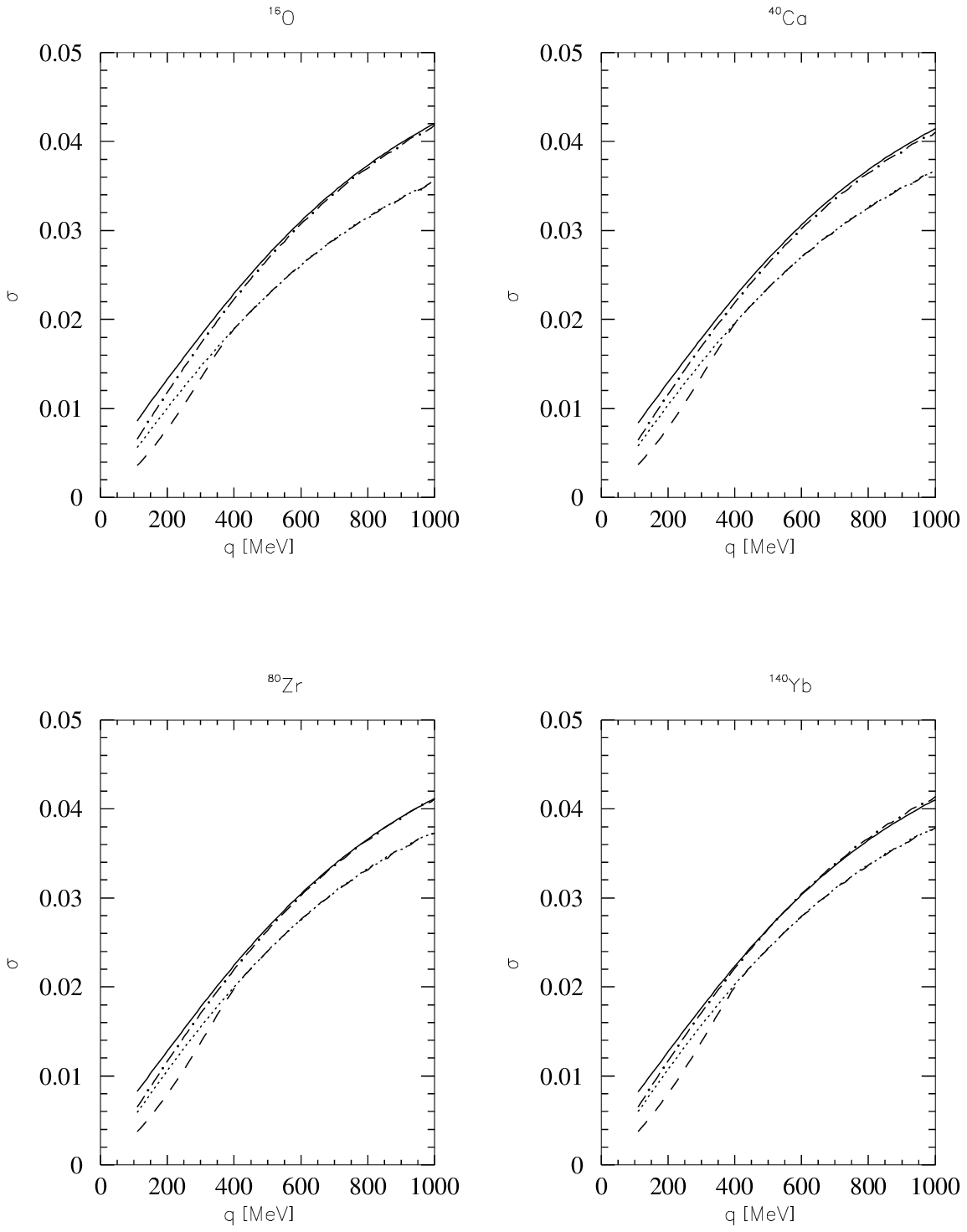}
\vskip-3cm\hskip3.2cm
\caption{\protect\label{Fig15}
The variance $\sigma$.
Dashed lines: RFG, with $\k$ evaluated according to eq.~(\protect\ref{IV.20});
dotted lines: RFG, with $\k$ evaluated according to 
eq.~(\protect\ref{IV.20}) and the energy shift suggested in 
eq.~(\protect\ref{XY2}); dash-dotted lines: RFG, with the energy shift in 
eq.~(\protect\ref{XY2}) as well as with the scaled Fermi momentum given 
by eq.~(\protect\ref{kfbarp}); solid lines: hybrid model.
}
\end{figure}

Both the upward shift and the broadening of the response are seen even
more clearly in figs. \ref{Fig13}, \ref{Fig14} and \ref{Fig15}
where the normalized sum rule, energy-weighted
sum rule (EWSR) and variance $\sigma$ of the HM and RFG are
displayed. 

With respect to the Coulomb sum rule of the HM shown in fig.~\ref{Fig13} we 
first notice that a small
deviation from one at large q might be expected, since a low-density 
approximation was employed in deriving the reduction factor, although such a 
deviation never exceeds 1-2\%. 

Second, to help in appreciating the role of 
confinement, we display together with the HM results three different versions
of the RFG sum rule. Thus in the figure the solid lines correspond to 
the case in which the integration is performed from $\omega=0$ up to the light 
cone (non-Pauli-blocked RFG), while for the dotted lines the integration 
starts from the RFG separation energy. In this last instance the sum rule
receives contributions even from negative values of $\omega$, since in fact 
the threshold energy in the RFG is negative (namely $\ E_S$). 

~~

Finally the 
dash-dotted lines correspond to the RFG sum rule with the Pauli principle 
included and the integration ranging from $\omega$ = 0 up to the light cone. 
A comparison of the dotted and solid lines (hybrid model) shows the 
impact of the confinement, insofar as the former rises vertically at 
threshold, while confinement smooths out the approach to the asymptotic 
sum rule. This behaviour is further softened by the action of the Pauli 
principle, as reflected in the dash-dotted lines.

\begin{table}
 \hskip-1cm
 \footnotesize{
 \begin{tabular}{||c||c|c|c|c||c||c|c|c|c||}
 \hline\hline
 ~&\multicolumn{4}{c||}{$E_S=-T_F$}&
 ~&\multicolumn{4}{c||}{$E_S=8$ MeV}\\
 \hline\hline
 $q$ (GeV/c)&0.5 &1&1.5&2&
 $q$ (GeV/c)&0.5 &1&1.5&2\\
 \hline
  $^{16}$O&  0.02& -0.86& -1.43& -1.31& $^{16}$O& -0.35& -1.39& -2.13& -2.10\\
 \hline
  $^{40}$Ca& -0.03& -0.34& -0.87& -1.19& $^{40}$Ca& -0.38& -0.88& -1.58& -2.00\\
 \hline
  $^{80}$Zr&  0.02& -0.12& -0.56& -1.11& $^{80}$Zr& -0.33& -0.67& -1.29& -1.94\\
 \hline
 $^{140}$Yb&  0.08& -0.16& -0.36& -1.06&$^{140}$Yb& -0.26& -0.72& -1.10& -1.90\\
 \hline
  $^{224}?$&  0.15& -0.10& -0.22& -0.63& $^{224}?$& -0.19& -0.67& -0.97& -1.48\\
 \hline
  $^{336}?$&  0.21& -0.06& -0.12& -0.31& $^{336}?$& -0.13& -0.64& -0.88& -1.17\\
 \hline\hline\hline
 ~&\multicolumn{4}{c||}{$E_S=30$ MeV}&
 ~&\multicolumn{4}{c||}{$E_S=50$ MeV}\\
 \hline\hline
 $q$ (GeV/c)&0.5 &1&1.5&2&
 $q$ (GeV/c)&0.5 &1&1.5&2\\
 \hline
  $^{16}$O& -0.73& -1.90& -2.74& -2.75& $^{16}$O& -1.22& -2.45& -3.36& -3.40\\
 \hline
  $^{40}$Ca& -0.75& -1.37& -2.17& -2.64& $^{40}$Ca& -1.22& -1.91& -2.78& -3.27\\
 \hline
  $^{80}$Zr& -0.69& -1.16& -1.87& -2.57& $^{80}$Zr& -1.15& -1.69& -2.47& -3.19\\
 \hline
 $^{140}$Yb& -0.62& -1.20& -1.68& -2.52&$^{140}$Yb& -1.08& -1.73& -2.28& -3.14\\
 \hline
  $^{224}?$& -0.55& -1.15& -1.55& -2.11& $^{224}?$& -1.01& -1.68& -2.15& -2.73\\
 \hline
  $^{336}?$& -0.49& -1.12& -1.46& -1.79& $^{336}?$& -0.95& -1.65& -2.06& -2.41\\
 \hline\hline
 \end{tabular}
  }
\vskip0.5cm
\caption{\protect\label{table2} The difference between left- and right-hand 
sides of eq.~(\protect\ref{mmmm}) in MeV. The RFG is taken with $\k$ defined by 
eq.~(\protect\ref{IV.20}).}
\end{table}

An interesting relationship can be established with respect to the EWSR shown 
in fig.~\ref{Fig14}: the EWSR quantifies the location of the maximum of the 
response, which in turn, as we have seen, occurs in the proximity of the 
energy where the scaling variable vanishes. On the basis of eq.~(\ref{XY2}) 
it is therefore natural to expect the following relationship to hold for the
energy-weighted sum-rule difference (EWSRD):
\begin{equation}
2m_N\left[ {{\Xi_1^{{\rm{HM}}}} \over {\Xi_0^{{\rm{HM}}}}} \ -
\ {{\Xi_1^{{\rm{RFG}}}} \over {\Xi_0^{{\rm{RFG}}}}}\right]\ =\ T_F +E_S.
\label{mmmm}
\end{equation}

Indeed in table \ref{table2}, where $k_F$ is chosen in accord with 
eq.~(\ref{IV.20}), one sees eq.~(\ref{mmmm})
to be very well obeyed for
a number of nuclei, over a large span of momentum transfers and for several
choices of the separation energy. Actually the departures from the 
predictions of eq.~(\ref{mmmm}), while mild, tend 
to grow with $q$, a symptom of their relativistic origin. However they stay very
small even for unreasonable large separation energies. 

With these ideas in mind we show as dotted curves in 
figs.~\ref{Fig10}--\ref{Fig12}, \ref{Fig14}, \ref{Fig15} the result 
of shifting the RFG answers by $E_S+T_F$, that is, using $\omega^\prime$ in 
eq.~(\ref{XY2}) to replace $\omega$ (this implies that $\lambda\rightarrow 
\lambda^\prime$ and $\tau\rightarrow\tau^\prime$, as discussed above, 
everywhere except in the single-nucleon form factors where $\tau$ is 
retained). Clearly this produces roughly the correct shift to yield 
agreement with the HM results. The EWSR shown as a dotted curve in 
fig.~\ref{Fig14} now approximates the HM answer, whereas the variance 
in fig.~\ref{Fig15} is largely unaffected by the shift.

This last observation suggests the strategy of modifying the Fermi 
momentum from the results given by eq.~(\ref{IV.20}) with the hope of 
adjusting the variance so that the ``renormalized'' RFG model has the 
same value of $\sigma$ as the HM and yet at the same time has the shift 
discussed above. Since the EWSR and variance are primarily affected by the 
shift and scaled Fermi momentum, respectively, in an almost ``orthogonal'' 
way, we anticipate being able to obtain agreement for all of the first 
three energy-weighted moments. Thus, in figs. \ref{Fig10}--\ref{Fig12}, 
\ref{Fig14}, \ref{Fig15} we also shown as dash-dotted curves results 
where the shift by $E_S+T_F$ discussed above is performed and as well 
where the Fermi momentum is scaled according to
\begin{equation}
{\overline k_F}^\prime\equiv \alpha_F {\overline k_F},
\label{kfbarp}
\end{equation}
where $\alpha_F$ is the dimensionless factor required to make this 
renormalized RFG model have the same variance as the HM.  In table~\ref{table1} 
we give the values of $\alpha_F$ required to accomplish this; they are found 
to obey the relationship
\begin{equation}
\alpha_F= 1.08 + \frac{1.6}{A}
\label{alphaf}
\end{equation}
reasonably well. The resulting values of ${\overline k_F}^\prime$ are 
shown in fig.~\ref{Fig9} where it is clear that the trend is in the right 
direction, the data now falling only about 10\% higher than the scaled 
Fermi momenta for heavy nuclei.

We deduce from this that the ``Fermi momentum'' one uses in other applications 
of the RFG should be viewed with some caution. If only properties of the 
$A$-body nuclear ground state and of low-lying states in the $A-1$ daughter 
nucleus are used to fix the effective Fermi momentum (as in the present 
work when spectral functions of RFG and HM are compared), then one value 
emerges, namely ${\overline k_F}$. In contrast, when cognizance is taken of 
the full $A$-body final nuclear state with a nucleon far up in the 
continuum by using the QE responses in a model-to-model comparison, then a 
larger effective Fermi momentum, ${\overline k_F}^\prime$, is found. Even 
extrapolating to $A=\infty$ still produces an 8\% upward scaling.

\section{Conclusions\label{sect7}}

In the present work we have achieved the goals set out in the Introduction. 
First, we have succeeded in deriving the non-Pauli-blocked RFG directly 
from the PWIA by constructing covariant initial and final (non-interacting) 
nuclear states, and from these obtaining the RFG spectral function. We have 
used the deepened understanding that comes from this approach to re-examine 
the issues of defining a reduced response function with its 
attendant properties, $y$-scaling, $\psi$-scaling and energy-weighted 
moments including the zeroth moment or Coulomb sum rule.

The second general goal of the present work was to develop a simple, tractable 
model for the nuclear spectral function that incorporates the confinement of 
nucleons in the initial state. Our specific model, a hybrid model with 
harmonic oscillator wave functions, is designed in a way that facilitates 
taking the $A\rightarrow\infty$ with minimal numerical effort. From 
comparisons between the HM and RFG spectral functions it is possible to 
define an average Fermi momentum ${\overline k_F}$ and to examine its 
behaviour in the large-$A$ limit. Typically the value obtained falls 
below that extracted from experimental determinations using QE scattering by 
roughly 20\%, and only for the innermost shells of very heavy nuclei is 
the standardly accepted nuclear matter Fermi momentum reached. 

Further comparisons of HM and RFG through reduced response functions and 
the resulting energy-weighted sum rules tell us that the global nature 
of the spectral functions is quite similar for the two models, although 
the detailed distribution in missing energy and missing momentum is not. 
As a consequence, while inclusive lepton scattering responses turn out to 
be similar, for exclusive scattering the models would yield radically 
different results, as expected. In particular, the definition of the 
reduced inclusive longitudinal response requires a reduction factor $H_L$ 
which can be model dependent and so rend the various energy-weighted 
sum rules, including importantly the Coulomb sum rule, model dependent. 
We find, however, that $H_L^{\rm RFG}$ and $H_L^{\rm HM}$ are only very 
weakly model dependent. The former proceeds from the on-shell single-nucleon 
current whereas the latter starts from off-shell prescriptions for the 
single-nucleon current, and yet, when the differing kinematics in the 
two models are correctly incorporated, the final results are almost 
indistinguishable. This bodes well for studies of scaling and energy-weighted 
moments for then the (uncontrollable) model dependence will be weak and it 
should suffice for models that are not too dissimilar from the HM to use 
$H_L^{\rm RFG}$. Such a procedure has, for example, been followed in 
recent experimental re-analyses of the Coulomb sum rule.

The further understanding of the EWSR coming from the model-to-model 
comparisons made in the present work also suggest refinements. Specifically, 
when examining the first moments of the HM and RFG we find that the responses 
are offset by $E_S+T_F$. Shifting the energy transfer by this amount yields 
the same first moment in the two models. Furthermore, comparing the variances 
in the two models we find the need for a larger effective Fermi momentum 
in the RFG to yield agreement with the HM result, which also produces a 
tendency in the right direction for agreement with experiment.  We conclude 
from these model-to-model comparisons that, with an energy-shift and rescaled 
Fermi momentum whose origins are now better understood than previously, an 
effective RFG response can be obtained whose first three energy-weighted 
moments agree quite well with the HM developed in this work. In the course
of reaching this deeper understanding of the two models we have also 
sharpened our knowledge about the $y$-scaling, $\psi$-scaling, 
$\psi^\prime$-scaling and Coulomb sum rule properties of quasielastic 
responses.

\newpage
\centerline{\large\bf Acknowledgements}

~

It is a pleasure to thank Mrs L. Opisso for the careful typing of the 
manuscript. The authors also wish to acknowledge the support provided under 
the M.I.T./I.N.F.N. exchange program during the course of this work.

\end{document}